\documentclass[12pt,a4paper]{article}
\usepackage{amsfonts}
\usepackage{a4wide}
\usepackage{amssymb}
\usepackage{amsmath,cite}
\usepackage{bbm}
\usepackage{ifpdf}
\ifpdf
\usepackage[pdftex]{graphicx}
\DeclareGraphicsExtensions{.pdf}
\usepackage[pdftex,implicit]{hyperref}

\hypersetup{%
  pdftitle    = {Seven--branes Revisited},
  pdfkeywords = {Supersymmetry, superstrings, branes, BPS solutions},
  pdfauthor   = {E. Bergshoeff, J. Hartong, T. Ortin and D. Roest},
  pdfcreator  = {pdf\LaTeXe\ with package \flqq hyperref\frqq},
  pdfproducer = {pdf\LaTeXe\ with package \flqq hyperref\frqq},
  pdfpagemode = None,  
  pdffitwindow= true,  
  unicode     = true,
  plainpages  = true,
  colorlinks  = true,  
  citecolor   = blue,
  urlcolor    = red,
  linkcolor   = black
}
\newcommand{\hepth}[1]{arXiv:{\tt \href{http://www.arXiv.org/abs/hep-th/#1}{hep-th/#1}}}

\else
  \usepackage[dvips]{graphicx}
\DeclareGraphicsExtensions{.eps}
  \usepackage[unicode,implicit]{hyperref}
  \newcommand{\hepth}[1]{arXiv:{\tt hep-th/#1}}

\fi
\makeatletter \@addtoreset{equation}{section} \makeatother

\begin{document}

\begin{flushright}
\small
\texttt{hep-th/0612072}\\
UG-06-09\\
IFT-UAM/CSIC-06-51\\
\hfill UB-ECM-PF-06-34 \\
\normalsize
\end{flushright}

\begin{center}

\vspace{.7cm}

{\LARGE {\bf Seven--branes and Supersymmetry}} \\

\vspace{8mm}

\begin{center}
    Eric A. Bergshoeff$^{\diamondsuit}$, Jelle Hartong$^{\diamondsuit}$, Tom\'{a}s Ort\'{\i}n$^{\clubsuit}$ and
    Diederik Roest$^{\spadesuit}$ \\[5mm]
    {\small\slshape
    $\diamondsuit$ Centre for Theoretical Physics, University of Groningen, \\
    Nijenborgh 4,
    9747 AG Groningen, The Netherlands \\
    {\upshape\ttfamily E.A.Bergshoeff, J.Hartong@rug.nl} \\[2mm]
    $\clubsuit$ Instituto de F\'{i}sica Te\'{o}rica UAM/CSIC, \\ Facultad de
    Ciencias C-XVI, C.U. \\
    Cantoblanco, E-28049-Madrid, Spain \\
    {\upshape\ttfamily Tomas.Ortin@cern.ch} \\[2mm]
    $\spadesuit$ Departament Estructura i Constituents de la Materia, \\
    Facultat de F\'{i}sica, Universitat de Barcelona, \\
    Diagonal, 647, 08028 Barcelona, Spain \\
    {\upshape\ttfamily droest@ecm.ub.es}}
\end{center}

\vspace{8mm}

{\bf Abstract}

\end{center}

\begin{quotation}

  We re-investigate the construction of half-supersymmetric 7-brane solutions
  of IIB supergravity.  Our method is based on the requirement of having
  globally well--defined Killing spinors and the inclusion of
  $SL(2,\mathbb{Z})$--invariant source terms.  In addition to the well-known
  solutions going back to Greene, Shapere, Vafa and Yau we find new
  supersymmetric configurations, containing objects whose monodromies are not
  related to the monodromy of a D7-brane by an $SL(2,\mathbb{Z})$
  transformation.

\end{quotation}

\newpage
\pagestyle{plain}

\section{Introduction}

The construction of half-supersymmetric 7-brane solutions goes
back to the classic work \cite{Greene:1989ya} where these were
presented as cosmic string solutions of a $D=4$ gravity plus
dilaton-axion system. Later, after the invention of D-branes
\cite{Polchinski:1995mt}, these solutions were oxidized to $D=10$
dimensions and re-interpreted as D7-brane solutions
\cite{Gibbons:1995vg}. Since then, D7-branes, in particular in the
form of D3-D7-brane systems, have found important applications in
model building, see e.g.~\cite{Grana:2001xn,
Karch:2003nh,Kirsch:2005uy}, and cosmology, see
e.g.~\cite{Dasgupta:2002ew, Kachru:2003aw, Burgess:2003ic}.

The original motivation of \cite{Greene:1989ya} was not the
construction of cosmic string solutions as such, but the
investigation of supersymmetric String Theory backgrounds that are
more general than the direct product of 4-dimensional Minkowski
space--time and a Calabi-Yau 3-fold.  The construction of
\cite{Greene:1989ya} assumes that the gravity plus dilaton-axion
system arises from compactification on a torus $T^{2}$, the
complex \textit{axidilaton} field $\tau$ being its modular
parameter. The cosmic string solutions found in
\cite{Greene:1989ya} can be seen as products of a 2-dimensional
Minkowski space--time (the worldsheet of the cosmic string) and a
nontrivial 4-dimensional space with the structure of a $T^{2}$
fibration over a 2-dimensional base space (the transverse space of
the cosmic string). Supersymmetry requires this 4-dimensional
space to be a Calabi-Yau 2-fold. Within the context of $D=10$ IIB
supergravity one must for this purpose rely on a 12-dimensional
F-theory \cite{Vafa:1996xn}.

It is the purpose of this work to re-analyze half-supersymmetric
7-brane solutions of IIB supergravity without invoking a
higher-dimensional origin of the gravity plus axidilaton system.
We will, instead, directly analyze the IIB supergravity Killing
spinor equations, taking into account all their symmetries, and we
will require that there exists a globally well--defined Killing
spinor. We will find that this supersymmetry requirement is less
restrictive than the one imposed in \cite{Greene:1989ya}.

Another distinguishing feature of our analysis is that we add
$SL(2,\mathbb{Z})$-invariant source terms to the equations of
motion\footnote{Strictly speaking the source terms are
  $SL(2,\mathbb{Z})$--invariant provided we also transform the constants that
  occur in these source terms, see eq.~(\ref{actioncoupledsystem}).}. These
source terms represent the coupling of a 7-brane to the IIB
supergravity background.  They enable us to derive an expression
for the 7-brane solution in a neighborhood of the brane source as
well as the monodromy of the fields around the brane source in
terms of the brane source charges.

In general, D7-branes do not come alone since this leads to
singularities at a finite distance from the D7-brane. To obtain a
globally well--defined solution one can add other 7-brane objects
whose monodromy is not related to the monodromy of a D7-brane by
an $SL(2,\mathbb{Z})$-transformation\footnote{In
  this work we will also present an alternative way to cure the singularity by the
  addition of an $SL(2,\mathbb{Z})$-transformed D7-brane, see
  section \ref{subgroup}.}.  We will call these objects, for reasons that will
become clear soon, ``$\det\, Q > 0$-branes''. The existence of
these new objects may be anticipated from the observation that the
standard Ramond--Ramond (RR) potential $C_{(8)}$ that couples to
the D7-brane is part of a triplet $C_{(8)}^{\alpha\beta}$ of
8-form potentials under $SL(2,\mathbb{Z})$, i.e.

\begin{equation}
C_{(8)}\hskip .3truecm \rightarrow\hskip .3truecm
C_{(8)}^{\alpha\beta}\,,
\end{equation}

\noindent and that not each combination of potentials is related
to $C_{(8)}$ via an $SL(2,\mathbb{Z})$--trans\-formation.

The multiple D7-brane solutions of \cite{Greene:1989ya} can be
viewed as special configurations where the $\det\, Q > 0$-branes
occur in particular groups such that their masses and monodromies
cancel amongst each other and one is left with multiple D7-branes
only. We will present new half-supersymmetric configurations where
these cancellations do not occur and we will discuss the
properties of these solutions. A distinguishing feature of these
new solutions is that the metric has a deficit angle at the
position of each $\det\, Q > 0$-brane that is not cancelled. We
will construct special solutions that can be used as the basic
building blocks for constructing all multiple 7-brane solutions
including the multiple D7-branes of \cite{Greene:1989ya}.

It is not clear what the correct interpretation of the $\det\,
Q>0$-branes is within String Theory. Part of this work's
motivation was to show explicitly that within the context of
supergravity one can allow for such objects. At present their
world-volume dynamics is not well-understood.

This paper is organized as follows: In section \ref{sourceterms}
we will discuss the $SL(2,\mathbb{Z})$--invariant source terms.
The analysis of the equations of motion, with emphasis on the
existence of a Killing spinor, will be the subject of section
\ref{sec:eom}. We will discuss a BPS equation for the 7-brane
solutions in section \ref{sec:energy}. Next, we will describe how
to construct globally well--defined 7-brane solutions in section
\ref{sec:monodromy}.  Explicit examples of solutions, old ones and
new ones, will be presented in section \ref{sec:examples}.
Finally, we give our conclusions in section \ref{sec:conclusions}.


\section{Seven--brane Source Terms}\label{sourceterms}

The gravity plus axidilaton system we are going to consider is a
consistent truncation of the IIB supergravity action
\cite{SW,schwarz,HW} in which only the metric, the RR 0-form
(axion) $\chi$ and the dilaton $\phi$ are kept. These two scalar
fields appear in the complex combination $\tau=\chi+ie^{-\phi}$
(the \textit{axidilaton}), which parameterizes an
$SL(2,\mathbb{R})/SO(2)$ coset.

The coupling of a 7-brane, labelled by the real numbers $p,q,r$,
to the gravity plus axidilaton system is described by the
following Einstein-frame ``pseudo action''\footnote{The reason
that we call the action
  \eqref{actioncoupledsystem} a pseudo action will become clear shortly.}:

\begin{equation}
\label{actioncoupledsystem}
\begin{array}{rcl}
S & = & {\displaystyle\frac{g_{s}^{2}}{16\pi G_{N}^{(10)}}\int
d^{10}x\sqrt{-g}
\left[R-\frac{\partial_{\mu}\tau\partial^{\mu}\bar\tau}{2\left(\text{Im}\tau\right)^2}
\right. }
\\
& & \\
& & {\displaystyle \left. -\int_{\Sigma}
d^{8}\sigma\sqrt{-g_{(8)}}\frac{\delta(x-X(\sigma))}{\sqrt{-g}}
\frac{1}{\text{Im}\tau}\left(p+q\vert\tau\vert^2+r\frac{\tau+\bar\tau}{2}\right)\right]\,
. }
\end{array}
\end{equation}

\noindent The 7-brane world-volume, $\Sigma$, is parameterized in
the above action by $\{\sigma^{i}, i=0,1,\ldots,7\}$.  The metric on the
world-volume is $g_{(8)ij}$ which is the pull-back of the target-space
Einstein-frame metric $g_{\mu\nu}$.  The embedding coordinates of the brane
are denoted by $X^{\mu}(\sigma)$, and so the pull-back is given by

\begin{equation}
g_{(8)ij}(\sigma)=\frac{\partial
X^{\mu}}{\partial\sigma^{i}}\frac{\partial
        X^{\nu}}{\partial\sigma^{j}}g_{\mu\nu}(X)\,.
\end{equation}

\noindent We are only considering objects for which in the static
gauge the transverse scalars are set equal to zero, i.e.~we do not
consider fluctuations of the world-volume.  The source term in
\eqref{actioncoupledsystem} should be interpreted as adding a
purely static object to the theory. Note that the source term is
linear in $p,q$ and $r$. This is related to the fact that, unlike
e.g.~strings, all 7-branes have the same half-supersymmetry
projection operator, see \eqref{susyprojection}, which is
invariant under $SL(2,\mathbb{R})$ transformations.

In the coefficient in front of the above action $g_{s}$ is the string coupling
constant (i.e.~the vacuum expectation value of $e^{\phi}$ measured at
infinity) and $G^{(10)}_{N}$ is the 10-dimensional Newton constant which is
given by

\begin{equation}
G^{(10)}_{N}= 8\pi^{6}g_{s}^{2}\ell_{s}^{8}\, ,
\end{equation}

\noindent where $\ell_{s}$ is the string length
$\sqrt{\alpha^{\prime}}$.  Then, the coefficient in front of the
action \eqref{actioncoupledsystem} is independent\footnote{One may
recover the form of the action as it appears in low--energy
perturbative string theory which has a factor $g_{s}^{-2}$ in
front of it by going to the so-called modified Einstein-frame
metric $\tilde{g}_{\mu\nu}\equiv g_{s}^{1/2} g_{\mu\nu}$
\cite{Maldacena:1996ky}.} of $g_{s}$. The relative numerical
coefficient between the brane probe action and the bulk action is
just 1 due to the coincidence:

\begin{equation}
T_{D7}g_{s}=\frac{g_{s}^{2}}{16\pi G_{N}^{(10)}}\, .
\end{equation}

The bulk action is invariant under $SL(2,\mathbb{R})$ transformations which
act on the axidilaton according to

\begin{equation}
\tau\rightarrow\Lambda\tau \equiv \frac{a\tau+b}{c\tau+d}
\quad\text{where}\quad\Lambda= \left(
\begin{array}{cc}
a & b \\
c & d \\
\end{array}
\right)\in SL(2,\mathbb{R})\, ,
\end{equation}

\noindent and leave the Einstein-frame metric $g_{\mu\nu}$
invariant.  Observe that the coefficient in front of the action is
$SL(2,\mathbb{R})$-invariant precisely because it does not depend on $g_{s}$.

The worldvolume term in the action is also $SL(2,\mathbb{R})$-invariant
provided that the real constants $p,q,r$, arranged in the traceless matrix

\begin{equation}
\label{pqr} Q \equiv \left(
\begin{array}{cc}
r/2 & p \\
-q & -r/2 \\
\end{array}
\right)\, ,
\end{equation}

\noindent transform in the adjoint representation of
$SL(2,\mathbb{R})$. Note that the determinant of this matrix,

\begin{equation}
\det{Q}=qp-r^{2}/4\, ,
\end{equation}

\noindent is $SL(2,\mathbb{R})$-invariant and can be used as a
label to distinguish between different conjugacy classes.

It is well known that the classical invariance of this theory is
broken by quantum-mechanical effects such as charge quantization
to $SL(2,\mathbb{Z})$. Hence, from now on we will only consider
this group. Observe that, actually, $\tau$ transforms only under
the group $PSL(2,\mathbb{Z})=SL(2,\mathbb{Z})/\{\pm \mathbbm{1}\}$
since $-\mathbbm{1}$ leaves it invariant.

The reader may notice that the source term present in the pseudo
action \eqref{actioncoupledsystem} contains only a Nambu--Goto
(NG) term and no Wess--Zumino (WZ) term. At first sight this seems
surprising. For instance, in the case of the D7-brane, which
corresponds to the case that $p=1$ and $q=r=0$ the source term
contains only the dilaton and there is no source term for the
axion whereas the D7-brane is known to have a magnetic axionic
charge. The reason that we nevertheless will be able to reproduce
the D7-brane solution is that we will only consider solutions for
which the axidilaton $\tau$ is a holomorphic function of the two
coordinates transverse to the D7-brane. This input comes from a
consideration of the Killing spinor equations, see subsection
\eqref{sec:holonomy}. Since the dilaton and axion are combined in
one holomorphic function it is enough to consider a source term
for the dilaton only. The action \eqref{actioncoupledsystem} is
only a convenient tool for investigating supersymmetric 7-brane
solutions. That is the reason that we call it a pseudo action. For
the derivation of a proper action and a justification for the use
of action \eqref{actioncoupledsystem} we refer to
\cite{inprogress}.

$SL(2,\mathbb{Z})$-invariant 7-brane world-volume actions were
considered in \cite{Bergshoeff:2006ic} and were shown to preserve
half of the supersymmetries for all possible values of $p,q,r$.
The world-volume action describing a single D7-brane or any
$SL(2,\mathbb{Z})$ transform thereof has values $p,q,r$ which
satisfy the condition $-r^2/4+pq=0$, or $\det\, Q=0$. For this set
of 7-brane actions one can introduce a single Born-Infeld vector
in a target-space gauge-invariant and $SL(2,\mathbb{Z})$-invariant
manner \cite{Bergshoeff:2006gs}. This confirms the identification
of these objects as Dirichlet branes or $SL(2,\mathbb{Z})$
transforms thereof. It turns out (see section \ref{sec:monodromy})
that in constructing globally well--defined solutions containing a
D7-brane, objects with $\det\, Q>0$ play a crucial role while on
the other hand the possibility $\det\, Q<0$ never arises. It is
the purpose of this paper to find out more about the status of the
$\det\, Q>0$ objects.

By a ``D7-brane" we mean any representative element of the
$\text{det}\,Q=0$ $SL(2,\mathbb{R})$ conjugacy class\footnote{The
supergravity solutions in this article which describe the
space--time close to a 7-brane are characterized by the value of
$\text{det}\,Q$. This value labels $SL(2,\mathbb{R})$ conjugacy
classes. Whenever we speak of a conjugacy class we will always
mean of $SL(2,\mathbb{R})$ and not of $SL(2,\mathbb{Z})$.}. This
is because in constructing finite energy solutions we divide out
type IIB supergravity by the duality group $SL(2,\mathbb{Z})$ (or
a subgroup thereof). In doing so one can no longer distinguish the
various elements of a particular conjugacy class where each
conjugacy class is characterized by the value of $\text{det}\,Q$.


\section{The Equations of Motion}
\label{sec:eom}


\subsection{Supersymmetry and holonomy of the Killing spinor}
\label{sec:holonomy}

We are considering supersymmetric solutions of the system
\eqref{actioncoupledsystem} and we thus require that the following
Killing spinor equations are satisfied (using the supersymmetry
rules of  \cite{Bergshoeff:1995as}):

\begin{align}
\delta_{\epsilon}\lambda &
=\frac{i}{\tau-\bar\tau}\left(\gamma^{\mu}\partial_{\mu}\bar\tau\right)
\epsilon_{C}=0\,,
\label{susyvardilatino} \\
& \nonumber \\
\delta_{\epsilon}\psi_{\mu} &
=\left(\partial_{\mu}+{\textstyle\frac{1}{4}}\omega_{\mu}^{\;\;ab}\gamma_{ab}
+\frac{1}{4(\tau-\bar\tau)}\partial_{\mu}(\tau+\bar\tau)\right)\epsilon=0\,.
\label{susyvargravitino}
\end{align}

\noindent The Killing spinor $\epsilon$ can be written as
$\epsilon=\epsilon_1+i\epsilon_2$ where $\epsilon_1$ and $\epsilon_2$ are two
Majorana-Weyl spinors. The chirality of $\epsilon$ is negative, i.e.
$\gamma_{11}\epsilon=-\epsilon$. The $C$ operation leaves Majorana spinors
invariant. The equations \eqref{susyvardilatino} and \eqref{susyvargravitino}
transform covariantly under the following $SL(2,\mathbb{Z})$ transformations

\begin{align}
\label{SL(2,R)Killingspinor} &
\tau\rightarrow\frac{a\tau+b}{c\tau+d} \,, \quad
\lambda\rightarrow e^{3i\varphi}\lambda\,,\quad
\psi_{\mu}\rightarrow e^{i\varphi}\psi_{\mu}\,,\quad
\epsilon\rightarrow e^{i\varphi}\epsilon\,,\quad \left(
\begin{array}{cc}
a & b \\
c & d \\
\end{array}
\right)\in SL(2,\mathbb{Z})\, ,
\end{align}

\noindent where $\varphi = \textstyle{\frac{1}{2}}\text{arg}(c\tau+d)$. This means that $\epsilon$ transforms under the double
cover of $SL(2,\mathbb{Z})$. We define $-1=e^{i\pi}$ and $1=e^{i0}$. There is
no restriction on the range of $\varphi$.  Numbers such as $e^{-i\pi}$ and
$e^{2i\pi}$ lie on another Riemann sheet.

The discrete group $SL(2,\mathbb{Z})$ is generated by the two elements $T$ and
$S$, which are defined as follows:

\begin{equation}
    T=\left(\begin{array}{cc}
            1 & 1 \\
            0 & 1
            \end{array}\right)\,, \qquad
    S=\left(\begin{array}{cc}
            0 & -1 \\
            1 & 0
            \end{array}\right)\,.
\label{SandT}
\end{equation}

\noindent Observe that, unlike $\tau$, the Killing spinor
$\epsilon$ does transform under $S^2 = -\mathbbm{1}$ as $\epsilon
\rightarrow i \,\epsilon$. Under $S^4=\mathbbm{1}$ we have
$\epsilon\rightarrow -\epsilon$. Only $S^8$ acts as the identity
on $\epsilon$.

The transformation rules of the spinors tell us that they carry $U(1)$ charge.
Actually, the term

\begin{equation}
\mathcal{Q}_{\mu}\equiv {\textstyle\frac{1}{2i}}
\frac{\partial_{\mu}(\tau+\bar\tau)}{(\tau-\bar\tau)}\, ,
\end{equation}

\noindent in the gravitino supersymmetry transformation rule
\eqref{susyvargravitino} is a $U(1)$ connection and the whole operator that
acts on $\epsilon$ is a $U(1)$ and Lorentz covariant derivative. In fact, the
coset $SL(2,\mathbb{R})/U(1)$ is a special K\"ahler manifold with K\"ahler
potential $\mathcal{K}=\log{\text{Im}\, \tau}$ and the above $U(1)$ connection
is nothing but the pullback over the spacetime of the K\"ahler connection

\begin{equation}
\mathcal{Q}= {\textstyle\frac{1}{2i}} (d\tau
\partial_{\tau}\mathcal{K} -d\bar{\tau}
\partial_{\bar{\tau}}\mathcal{K})=
{\textstyle\frac{1}{2i}}\frac{d(\tau+\bar{\tau})}{(\tau-\bar{\tau})}\,
.
\end{equation}

\noindent Under isometries of the K\"ahler manifold (here the
group $SL(2,\mathbb{R})$) the K\"ahler potential is only invariant
up to K\"ahler transformations which become $U(1)$ transformations
of the K\"ahler connection.  This point will play a role in
discussing under which condition the Killing spinors are well
defined. We will also need the expression of the K\"ahler 2-form
$\Omega$, which is the field strength of the K\"ahler connection:

\begin{equation}
\label{eq:Kahler2form} \Omega\equiv d\mathcal{Q} =
{\textstyle\frac{1}{2i}}\frac{d\tau\wedge
d\bar{\tau}}{(\text{Im}\,\tau)^{2}}\, .
\end{equation}

The $SL(2,\mathbb{Z})$-invariant supersymmetry projection operator of a
7-brane extended in the directions $x^{1},\cdots,x^{7}$ is given by

\begin{equation}\label{susyprojection}
    P\epsilon=\tfrac{1}{2}\left(1-i\gamma_{\underline{0}\ldots
    \underline{7}}\right)\epsilon=\tfrac{1}{2}\left(1+i\gamma_{\underline{8}}
\gamma_{\underline{9}}\right)\epsilon=0\;.
\end{equation}

\noindent It follows that
$\left(\gamma_{\underline{8}}+i\gamma_{\underline{9}}\right)\epsilon=0=
\left(\gamma_{\underline{8}}-i\gamma_{\underline{9}}\right)\epsilon_C$.  If we
assume that $\tau$ and the metric do not depend on the worldvolume coordinates
$x^{0},\cdots,x^{7}$ and choose a conformally-flat transverse metric, then
Eq.~\eqref{susyvardilatino} tells us that
$\left(\partial_{8}-i\partial_{9}\right)\bar\tau=0$.  We define the complex
transverse coordinate $z=x^{8}+ix^{9}$ so that we now have
$\partial_{z}\bar\tau=0$, that is, $\tau$ is a holomorphic function. In
complex coordinates the condition on $\epsilon$ can be written as
$\gamma_{{\underline{z}}^{*}}\epsilon=0$. Under these conditions, the most
general 7-brane solution to equations \eqref{susyvardilatino} and
\eqref{susyvargravitino} is given by
\cite{Gibbons:1995vg,Meessen:1998qm,Lozano-Tellechea:2000mc,Bergshoeff:2002mb}

\begin{eqnarray}
ds^2 & = & -dt^2+d{{\vec x}_{7}}^{\;2}+(\text{Im}\tau)\vert
f\vert^2 dz d\bar
z\,,\label{metric}\\
& & \nonumber \\
\tau & = & \tau(z)\,,\quad f=f(z)\,,\label{tau}\\
&  & \nonumber \\
\epsilon & = & \left(f/\bar f\right)^{\!\!1/4}\!\!\epsilon_0\,,
\label{Killingspinor}
\end{eqnarray}

\noindent where $\epsilon_{0}$ is a constant spinor which
satisfies $\gamma_{{\underline{z}}^{*}}\epsilon_{0}=0$. We will
study in the next sections how these supersymmetric configurations
solve the classical equations of motion corresponding to the
action \eqref{actioncoupledsystem}.

The functions $\tau$ and $f$ are assumed to be defined on the Riemann sphere.
The form of the solution is therefore fixed up to $SL(2,\mathbb{C})$
transformations

\begin{equation}\label{globalcoordinatefreedom}
z\rightarrow\frac{az+b}{cz+d}\,, \hspace{1cm} \left(
\begin{array}{cc}
a & b \\
c & d \\
\end{array}
\right)\in SL(2,\mathbb{C})\, .
\end{equation}

\noindent These are the most general global coordinate
transformations that do not change the structure of the branch cuts and
singularities of $\tau$ and $f$ in the complex $z$-plane.  Note that locally
(but not globally) we can always choose a basis in which $f(z)=1$.

Although the configurations \eqref{metric}--\eqref{Killingspinor}
are locally supersymmetric, they must satisfy further conditions
to be globally well--defined and supersymmetric. The main issue
here will be the possible multi-valuedness of $\tau(z)$ and
$f(z)$, which in general will be holomorphic functions with
singularities and branch cuts and which appear in the metric and
Killing spinor.

The axidilaton $\tau$, being a physical field of IIB supergravity, must be
single-valued. However, when constructing solutions, we will consider the IIB
supergravity theory divided out by (a subgroup of)
$SL(2,\mathbb{Z})$\footnote{In the case of dividing out by $SL(2,\mathbb{Z})$
  this means that we are effectively dealing with a 64+64 $N=1, D=8$
  supergravity multiplet coupled to a 8+8 vector multiplet instead of the
  128+128 IIB supergravity multiplet. Further, there are additional vector
  multiplets coming from the presence of 7-branes. The reduction is over the
  two directions transverse to the 7-branes and is triggered by the fact
  that $\tau$ is not an arbitrary holomorphic function of $z$.\label{dividingbySL2Z}}. Therefore, we
consider values of $\tau$ related by transformations belonging to
(a subgroup of) $PSL(2,\mathbb{Z})$ as equivalent. In particular,
$\tau(z)$ may jump to $\Lambda\tau(z)$ when crossing a branch cut.
In other words, it may have a non-trivial monodromy contained in
(a subgroup of) $PSL(2,\mathbb{Z})$.

We will now derive the transformation rule for the function $f$
when going around a 7-brane by requiring that the holonomy of
$\epsilon$ be well--defined. The holonomy of the Killing spinor is
computed with respect to the generalized connection in
\eqref{susyvargravitino}, which is the sum of the Lorentz
connection and $U(1)$ connection. The integrability condition of
\eqref{susyvargravitino} requires that the total curvature
vanishes but the Riemann curvature of the transverse space and the
$U(1)$ curvature are, separately, non-trivial.

If we parallel-transport the Killing spinor $\epsilon$ using the connection in
\eqref{susyvargravitino}, evaluated on the solution (\ref{metric},\ref{tau})
from a base point $b$ around a closed loop $\gamma_{b}$ it can be shown that
the holonomy (with respect to the Lorentz group) of $\epsilon$ is given by

\begin{equation}
\label{holonomy} \epsilon(b) \rightarrow
\exp\left({\textstyle\frac{i}{2}}\,\text{Im}
\oint_{\gamma_{b}}(\log{f})^{\prime}dz\right)\epsilon(b)\, ,
\end{equation}

\noindent where the prime denotes differentiation with respect to
$z$.

The holonomy phase factor will depend on the base point $b$ but only on the
homotopy class of $\gamma_{b}$ due to the vanishing total curvature. We
require\footnote{In general one can also allow for nontrivial spin structures
  but we will not do so here.} it to be an $SL(2,\mathbb{Z})$ transformation
as given in equation \eqref{SL(2,R)Killingspinor}

\begin{equation}\label{holonomycondition}
\exp\left({\textstyle\frac{i}{2}}\,\text{Im}
\oint_{\gamma_{b}}(\log{f})^{\prime}dz\right)= e^{i\varphi}\, ,
\end{equation}

\noindent such that the holonomy with respect to the generalized
connection is trivial. Let $\gamma_{b}$ be parameterized by $\lambda\in
[0,1]$. Then

\begin{equation}
\exp\left({\textstyle\frac{i}{2}}\,\text{Im}
\oint_{\gamma_{b}}(\log{f})^{\prime}dz\right)=
\left(\frac{f(\lambda=1)}{\vert f(\lambda=1)\vert)}\right)^{1/2}
\left(\frac{\vert f(\lambda=0)\vert}{f(\lambda=0))}\right)^{1/2}\,
.
\end{equation}

\noindent The requirement \eqref{holonomycondition} then leads to
the following condition for the function $f$

\begin{equation}
\label{monodromytrafof} f(\lambda=1)=(c\tau+d)f(\lambda=0)\, .
\end{equation}

\noindent Thus, when crossing a branch cut at the point $z$ we
must have

\begin{equation}\label{monf}
f(z) \rightarrow (c\tau(z)+d)f(z)\, .
\end{equation}

\noindent For the convenience of the reader we summarize some of
the $SL(2,\mathbb{Z})$ properties of $\tau, f$ and $\epsilon$ in table
\ref{convenience}.

\begin{table}[h]
\begin{center}\hspace{-.5cm}
\begin{tabular}{||c|c|c||}
  \hline
    {\small fields} & {\small group} &  {\small order of $S$}\\
  \hline \hline
$\tau$&$PSL(2,\mathbb{Z})$&2\\
$f$&$SL(2,\mathbb{Z})$&4\\
$\epsilon$&double cover&8\\
  \hline
\end{tabular}
\caption{Some $SL(2,\mathbb{Z})$ properties of $\tau, f$ and
$\epsilon$. } \label{convenience}
\end{center}
\end{table}

The metric $g_{\mu\nu}$ is a physical field which must be single-valued modulo
coordinate transformations. On the other hand, $\text{Im}\, \tau$ appears
explicitly in the expression \eqref{metric} for $g_{\mu\nu}$ and it may
transform into $|c\tau+d|^{-2}\text{Im}\, \tau$ when crossing a branch cut. In
general, the extra factor $|c\tau+d|^{-2}$ cannot be eliminated by an
$SL(2,\mathbb{C})$ transformation and, thus, it must be compensated by $f(z)$.
From \eqref{monf} we see that the metric remains invariant when going around a
7-brane.

It is worth pausing a moment to compare the present situation with
that of Ref.~\cite{Howe:1995zm}. The system of Killing spinor
equations studied there is essentially identical to the system
studied here\footnote{The eight worldvolume dimensions of the
  7-brane solutions do not play any role and we can view this system as,
  effectively, $2+1$-dimensional. We will do this to compute the mass in
  section~\ref{sec:energy}.}. In particular, the gravitino supersymmetry
transformation rule in \cite{Howe:1995zm} contains the Lorentz
connection and a $U(1)$ connection which, on shell, is, up to
gauge transformations, the K\"ahler connection of the scalar
manifold to which a Chern-Simons supergravity is coupled. In fact,
they find cosmic-string solutions that include those studied here
and those found in \cite{Meessen:2006tu} in $N=2,D=4$ theories
with vector multiplets. The authors of \cite{Howe:1995zm},
however, required the Killing spinors to have trivial monodromies
up to signs corresponding to the non-trivial spin structure of the
transverse space while here it is required that the monodromies
should be trivial up to $U(1)$ transformations corresponding to
the non-trivial spin$^{c}$ structures of the transverse space.
These are the right structures for $U(1)$-charged spinors and they
allow for more general monodromies than those considered in
\cite{Howe:1995zm}.


\subsection{The scalar equations of motion}

In this section we are going to study how the supersymmetric configurations
found in the previous section solve the scalar equation of motion with sources
derived from the action \eqref{actioncoupledsystem}.  We perform a variation
of the action \eqref{actioncoupledsystem} with respect to $\bar\tau$ and use
the metric \eqref{metric}. This leads to the following equation of motion for
$\tau$:

\begin{equation}\label{eqfortau}
\partial\bar\partial\tau-2\frac{\partial\tau\bar\partial\tau}{\tau-\bar\tau}=
-{\textstyle\frac{i}{4}}\delta(z-z_{0},\bar z-\bar
z_{0})\left(p+q\tau^2+r\tau\right)\,.
\end{equation}

\noindent Due to the presence of the delta function\footnote{We
define
\begin{displaymath}
\int{\textstyle\frac{i}{2}}dz\wedge d\bar z\,\delta(z,\bar z)=1\;.
\end{displaymath}
} we cannot at this stage assume that $\tau$ is a globally holomorphic
function.  Equation \eqref{eqfortau} can be integrated as follows. Let $R$ be
an infinitesimal disk $\vert z-z_0\vert \leq\delta$ and let us denote its
boundary by $\gamma_{\delta}$.  Integrating equation (\ref{eqfortau}) over $R$
we obtain

\begin{equation}
\lim_{\delta\to 0}\int_{R}\left(\partial\bar\partial\tau
-2\frac{\partial\tau\bar\partial\tau}{\tau-\bar\tau}\right)
{\textstyle\frac{i}{2}}dz\wedge d\bar z=
-{\textstyle\frac{i}{4}}\lim_{\delta\to
0}\oint_{\gamma_{\delta}}\tau^{\prime}
dz=-{\textstyle\frac{i}{4}}\left(p+q\tau^2+r\tau\right)_{z=z_0}\,,
\end{equation}

\noindent where the prime denotes differentiation with respect to
$z$. We have used Green's theorem\footnote{In complex notation Green's theorem
  for any real-analytic function $F$ defined on $R/\{z_0\}$ reads
\begin{displaymath}
    \int_{R}\partial\bar\partial F{\textstyle\frac{i}{2}}dz\wedge d\bar
    z={\textstyle\frac{i}{4}}\left(\oint_{\partial R}\bar\partial F d\bar z
-\oint_{\partial R}\partial F dz\right)\, .
\end{displaymath}
} to relate the integral over $R$ to the integral over the boundary
$\gamma_{\delta}$ and the fact that $\bar\partial\tau=0$ over
$\gamma_{\delta}$.

Assuming that when $q,r\neq 0$ the limit $\lim_{z\to z_0}\tau$ exists one may
write

\begin{equation}
    2\pi i\tau(z_0)=\lim_{\delta\to 0}\oint_{\gamma_{\delta}}\frac{\tau}{z-z_0}dz\,.
\end{equation}

Therefore we have

\begin{equation}
\label{inttau} \lim_{\delta\to 0}\oint_{\gamma_{\delta}}\left(2\pi
i\tau^{\prime}
-p\frac{1}{z-z_0}-q\frac{\tau^2}{z-z_0}-r\frac{\tau}{z-z_0}\right)dz=0\,.
\end{equation}

\noindent This form of the scalar equations of motion is
convenient to derive an approximation of the possible solutions close to the
source terms at $z_0$. This derivation goes as follows. We assume that the
integrand of \eqref{inttau} is an analytic function without any poles in the
interior of $\gamma_{\delta}$. Then it admits in $R$ a Taylor expansion

\begin{equation}\label{differentialeqtau}
2\pi i\tau^{\prime}-p\frac{1}{z-z_0}-q\frac{\tau^2}{z-z_0}
-r\frac{\tau}{z-z_0}=\sum_{n=0}^{\infty}a_n(z-z_0)^n\, .
\end{equation}

\noindent In the limit $\vert z-z_0\vert\rightarrow 0$ the poles
on the left hand-side will dominate all the terms on the right hand-side. In
this approximation the right hand-side of \eqref{differentialeqtau} can be put
to zero, and we are left with the homogeneous version of equation
\eqref{differentialeqtau}, i.e.

\begin{equation}
\label{differentialeqtau2} 2\pi
i\tau^{\prime}-p\frac{1}{z-z_0}-q\frac{\tau^2}{z-z_0}
-r\frac{\tau}{z-z_0}=0\,.
\end{equation}

\noindent The solutions to \eqref{differentialeqtau2} are

\begin{alignat}{2}
    & e^{2\pi i\tau/p}=z-z_0\quad&& \text{for $\text{det}\,Q=0$ and $q=r=0$}\,, \label{taudetq=0}\\
&&&\nonumber\\
    & c\left(\frac{\tau-\tau_0}{\tau-\bar\tau_0}\right)^{\frac{\pi}
    {\sqrt{\text{det}\,Q}}}=z-z_0\qquad&& \text{for
    $\text{det}\,Q>0$ and $q\neq 0$} \label{taudetq>0}\,,
\end{alignat}

\noindent where $Q$ is the matrix defined in \eqref{pqr},

\begin{equation}
\tau_0=-\frac{r}{2q}+\frac{i}{\vert q\vert}\sqrt{\text{det}\,Q}\,
,
\end{equation}

\noindent and $c\neq 0$ is a constant. Any $PSL(2,\mathbb{Z})$
transformation of these solutions is again a solution, leading to solutions
for all possible values of $p,q$ and $r$.

All these solutions turn out to have a non-trivial monodromy, as we assumed in
the previous section. It is always given by

\begin{equation}\label{localmonodromytau}
\tau\rightarrow e^{Q}\tau \qquad\text{where}\qquad
e^Q=\cos(\sqrt{\text{det}\,Q})I+
\frac{\sin(\sqrt{\text{det}\,Q})}{\sqrt{\text{det}\,Q}}Q\,.
\end{equation}

\noindent This identifies $e^Q$ as the monodromy matrix and
establishes the relation between monodromy and the $p,q,r$ charges of the
sources.  The two eigenvalues of $e^Q$ are $e^{\pm i\sqrt{\text{det}\,Q}}$.
Note that $\tau_{0}$ is a fixed point of the monodromy matrix $e^Q$ when
$\text{det}\,Q >0$. Equation \eqref{differentialeqtau2} also has solutions
with $\text{det}\,Q <0$. We do not consider these solutions here because in
dividing out type IIB by $SL(2,\mathbb{Z})$ or any subgroup thereof, points
which are fixed points under $e^Q$ with $\text{det}\,Q <0$ do not appear.

The left hand-side of expressions \eqref{taudetq=0} and \eqref{taudetq>0} can
be recognized as expansions of modular functions \cite{Schoeneberg} around
fixed points of some modular group of transformations.  In sections 5 and 6 we
will discuss the full solutions to the scalar field equations \eqref{inttau}
in terms of such modular functions and incorporate the above solutions as
approximations around certain fixed points.


\subsection{The Einstein equations of motion}

Varying the action \eqref{actioncoupledsystem} with respect to the metric and
substituting equations \eqref{metric} and \eqref{tau} one finds that the
$z\bar z$ component of the Einstein equations is given by

\begin{equation}
\label{eqforf}
\partial\bar\partial\log\vert f\vert^2=
-{\textstyle\frac{1}{2}}\delta(z-z_{0},\bar z-\bar z_{0})
\frac{i}{\tau-\bar\tau}\left(p+q\vert\tau\vert^2+r\frac{\tau+\bar\tau}{2}\right)\,,
\end{equation}

\noindent where $\partial=\frac{\partial}{\partial z}$. All other
components of the Einstein equation are identically zero.

Integrating equation (\ref{eqforf}) over a disk $R$ which is bounded by
$\gamma_{\delta}$ as defined in section \ref{sec:holonomy} and using that
$\bar\partial f=0$ we obtain

\begin{equation}\label{intfoverR}
\lim_{\delta\to
0}\,\text{Im}\oint_{\gamma_{\delta}}(\log{f})^{\prime}dz
=-\frac{i}{\tau-\bar\tau}\left(p+q\vert\tau\vert^2
+r\frac{\tau+\bar\tau}{2}\right)_{z=z_0}\,.
\end{equation}

Using equations \eqref{taudetq=0} and \eqref{taudetq>0} we can write

\begin{equation}
\label{intf-intermsof-pqr2} \lim_{\delta\to
0}\,\text{Im}\oint_{\gamma_{\delta}}(\log{f})^{\prime}dz
=-\,\text{sign}(q)\,\sqrt{\text{det}\,Q}\,,
\end{equation}

\noindent where $e^Q$ is the monodromy matrix of $\tau$ measured
when going around the contour $\gamma_{\delta}$.

The orders of the zeros/poles of the function $f(z)$ at $z=z_0$ determines the
deficit angle $\delta$ at the location of the source. Let $\gamma_{\eta}$ be a
closed circular contour of radius $\eta$ which encircles the point $z_0$. Then
we have

\begin{equation}
\label{zerosf} \delta = -\frac{1}{i} \lim_{\eta\rightarrow
0}\oint_{\gamma_{\eta}}(\log f)^{\prime}dz\,.
\end{equation}

\noindent Combining eqs.~\eqref{intf-intermsof-pqr2} and
\eqref{zerosf} we thus find the following expression for the deficit angle at
the location of the source:

\begin{equation}
\label{eq:deficit} \delta =  \text{sign}
(q)\sqrt{\text{det}\,Q}\,.
\end{equation}

\noindent Note that there is no deficit angle at the position of a
$\text{det}\,Q=0$-brane.

We will now derive an expression for the mass of the 7-brane
solution in terms of a bulk contribution and the deficit angles at
the position of the $\text{det}\,Q>0$ objects. For this purpose we
consider the 7-branes to be wrapped on a $T^{7}$ with radii
$R^{1},\cdots,R^{7}$ so they can be viewed as point-particles
moving in a $1+2$-dimensional space--time.

In general, the total energy of a massive particle in $1+2$ dimensions is
measured by the deficit angle at infinity via the formula \cite{Deser:1983tn}

\begin{equation}
m =\frac{1}{16\pi  G_{N}^{(3)}}\int d^{2}x\sqrt{|\gamma|}\,
R(\gamma)\,,
\end{equation}

\noindent where $ G_{N}^{(3)}$ is the (2+1)-dimensional Newton's
constant, related to the 10-dimensional one by

\begin{equation}
 G_{N}^{(3)}=   \frac{G_{N}^{(10)}}{(2\pi)^{7}R^{1}\cdots R^{7}}\, ,
\end{equation}

\noindent and $\gamma$ is the metric of the transverse space.

For static solutions in 2+1 dimensions one has
$G_{\;\;0}^{0}=-\frac{1}{2}R(\gamma)$, where $R(\gamma)$ is the Ricci scalar
of the metric $\gamma$.  We have $G_{\;\;0}^{0}=\frac{1}{2} T_{\;\;0}^{0}$ and
hence the energy is given by

\begin{equation}
\label{energydef} m=\frac{1}{16\pi  G_{N}^{(3)}} \int
d^{2}x\sqrt{|\gamma|}\, R(\gamma)=-\frac{1}{16\pi G_{N}^{(3)}}\int
{\textstyle\frac{i}{2}}dz\wedge d\bar
z\sqrt{|\gamma|}T^{0}_{\;\;0}\, .
\end{equation}

\noindent We have

\begin{equation}
\label{energymomentum} T^{0}_{\;\;0}=
-\frac{1}{\sqrt{|\gamma|}}\frac{1}{(\text{Im}\tau)^2}\partial\tau\bar\partial\bar\tau-\sum_{n}\frac{1}{\sqrt{|\gamma|}}\delta(z-z_n,\bar
z-\bar z_n)\frac{1}
    {\text{Im}\tau}\left(p+q\vert\tau\vert^2+r\frac{\tau+\bar\tau}{2}\right)\,,
\end{equation}

\noindent where $n$ labels the points $z_n$ where the particles
are located.  Using eqs.~\eqref{intfoverR},
\eqref{intf-intermsof-pqr2} and \eqref{eq:deficit} we obtain the
following expression for the energy:

\begin{equation}
\label{energy} m=\frac{1}{16\pi
G_{N}^{(3)}}\left(\int{\textstyle\frac{i}{2}}dz\wedge d\bar z
    \frac{\partial\tau\bar\partial\bar\tau}{(\mbox{Im}\tau)^{2}}+
2\,\sum_{n}\, \delta_n \right)\, ,
\end{equation}

\noindent where $\delta_{n}$ is the deficit angle at the location
of the $n$th particle at the point $z_{n}$.  Branes with
$\text{det}\,Q=0$ do not contribute to the sum in \eqref{energy}.
Their energy is solely given by the bulk contribution. A further
discussion of this can be found in section \ref{sec:monodromy}.


\section{The BPS Equation}
\label{sec:energy}

In this section we will derive a BPS equation for 7-brane solutions relating
the energy $m$ to the monodromy of an image object located at the asymptotic
region of the transverse space.  An image object \cite{Deser:1983tn} is an
unphysical object which one adds to the solution in order to identify the
transverse space with a sphere (one point compactification).  By consistency,
its charges (including the mass) must ``neutralize'' the solution.  In this
case, this means that the monodromy of the image object, which we will denote
by $e^{Q_{\infty}}$, is the inverse of the total monodromy measured when going
around all other objects. In addition its mass is such that the total deficit
angle adds up to $4 \pi$ (the transverse space has become a sphere).

The asymptotic region of the transverse space, the region $\vert
z\vert\rightarrow\infty$, corresponds to a single point on the Riemann sphere,
the point $z=\infty$.  This particular point on the Riemann sphere is the
location of the image object\footnote{Note that the location of the image
  object is arbitrary. We could have chosen to place it at any other point,
  $z_0$ say, of the Riemann sphere.  Generally speaking a point $z_0$ on the
  Riemann sphere is mapped to an asymptotic region of the transverse space
  when the physical distance from $z_0$ to any other point diverges as $\vert
  z\vert^{1-4 G_{N}^{(3)}m}$ while $m<1/4 G_{N}^{(3)}$ \cite{Deser:1983tn}.\label{asymptoticregion}}.
The asymptotic expansion of the metric \eqref{metric} generally
takes the form

\begin{equation}\label{asymptoticmetric}
    ds^{2}_{\infty}=-dt^2+d{{\vec x}_{7}}^{\;2}
+\text{cst}\vert z\vert^{-8G_{N}^{(3)}m} dz d\bar z\,,
\end{equation}

\noindent where $\tau_{\infty}=\text{cst}$ is the asymptotic value
of $\tau$ and where $f\rightarrow z^{-4G_{3}m}$ near $z=\infty$
with $m\ge 0$ the total mass of the solution. The space is
asymptotically conical with deficit angle $\delta = 8\pi
G_{N}^{(3)} m$. Not all contributions in \eqref{energy} to the
total mass $m$ need to be positive.  It is a special property of
(2+1)-dimensional space--times that one can allow for negative
deficit angles (negative point masses). In terms of the function
$f$ this statement is no other than saying that $f$ can have both
zeros (positive deficit angle) and poles (negative deficit angle).
We restrict to solutions for which the total mass $m\ge 0$.

From equation \eqref{asymptoticmetric} it is clear that the point $z=\infty$
must be a zero of the function $f$. The energy \eqref{energy} can be computed
as follows

\begin{equation}
m=\frac{1}{8\pi  G_{N}^{(3)}}\text{Im}\oint_{z=\infty}(\log
f)^{\prime}dz\,,
\end{equation}

\noindent where the contour integral encircles the point
$z=\infty$ (in a counter-clockwise direction). We must have, in order that
$\epsilon$ transforms correctly when going around infinity, that

\begin{equation}
\label{BPSbound} m=\frac{1}{8\pi  G_{N}^{(3)}}\text{Im}
\oint_{z=\infty}(\log{f})^{\prime}dz= -\,\text{sign}(q_{\infty})\,
\frac{\sqrt{\text{det}\,Q_{\infty}}}{8\pi  G_{N}^{(3)}}=
-\frac{\delta_{\infty}}{8\pi  G_{N}^{(3)}}\, ,
\end{equation}

\noindent where $Q_{\infty}$ is the charge matrix of the monodromy
of $\tau$ when going around $z=\infty$. Equation \eqref{BPSbound} may be
referred to as the BPS identity for 7-brane solutions. Note that, in order
that $m>0$ we must have $\delta_\infty<0$.

It is instructive to compare the BPS identity \eqref{BPSbound} with the mass
formula \eqref{energy}. The identification of the two formulae implies

\begin{equation}
\int \Omega = -2\,\sum_{n^{\prime}}\, \delta_{n^{\prime}}\, ,
\end{equation}

\noindent where the sum is extended to all ``holes'' in the
transverse space, including the asymptotic region and the l.h.s.~is the
integral of the pull-back of the K\"ahler 2-form defined in
\eqref{eq:Kahler2form}. This, in turn, implies that the deficit angles at the
locations of the particles/7-branes in transverse space, including the image
particle, can also be computed via line integrals of the pull-back of the
K\"ahler connection around the locations of the particles/7-branes in
transverse space

\begin{equation}
\label{eq:U1charge} \delta= {\textstyle\frac{1}{2}}\oint
\mathcal{Q}\, .
\end{equation}

\noindent This expression can be taken as the definition of $U(1)$
charge (see \cite{Howe:1995zm}). If we apply this formula to
compute the $U(1)$ charge at infinity \eqref{BPSbound} becomes a
relation between the mass and the $U(1)$ charge with the
characteristic form of a (saturated) BPS bound. It is now not too
difficult to see that the proof of \cite{Howe:1995zm} that all the
solutions of the system under consideration which have the
asymptotic behavior that allows to define mass and $U(1)$ charge
are automatically supersymmetric and both are related by the
saturated BPS bound which becomes an identity. In other words,
there are no ``black'' 7-brane solutions with a horizon.

Finally, observe that \eqref{eq:U1charge} can be written in the form

\begin{equation}
\delta = {\textstyle\frac{1}{2}} \text{Im}\, \oint
(\log{\text{Im}\, \tau})^{\prime}\, ,
\end{equation}

\noindent and, comparing with \eqref{zerosf} we see that the fact
that $\delta$ can be computed using either $\text{Im}\, \tau$ or
$f$ is a consequence of both functions having related monodromies,
which is something we required in order to have well--defined
Killing spinors and supersymmetry.


\section{Constructing Solutions}\label{sec:monodromy}

In this section we will discuss how to construct globally
well--defined solutions. The 7-brane configuration \eqref{metric},
\eqref{tau} contains two undetermined holomorphic functions:
$\tau(z)$ and $f(z)$. Both functions are defined on the Riemann
sphere, $\hat{\mathbb{C}}$.

An important role in constructing a globally well--defined
solution is the choice of the monodromy group which we will
discuss first.  Consider an arbitrary point $b\in\hat{\mathbb{C}}$
and form all possible closed loops with $b$ as their common base
point. The set of all monodromies measured when going around each
of these loops forms a group, called the monodromy group.  The
function $\tau$ transforms under $PSL(2,\mathbb{Z})$ and the
function $f$ transforms under $SL(2,\mathbb{Z})$. By monodromy
group we will always mean the monodromy group of $\tau$.

Consider a single D7-brane. This corresponds to $p=1$ in equation
\eqref{taudetq=0}. The monodromy of $\tau$ measured when going around a single
D7-brane is $\tau\rightarrow T\tau\equiv\tau+1$, where $T$ is defined in
\eqref{SandT}. Further, from equation \eqref{monf} it follows that
$f\rightarrow f$. The element $T\in PSL(2,\mathbb{Z})$ is of infinite order.
However, solutions containing only one object with this monodromy will have
infinite mass per volume element \cite{Greene:1989ya}. This is related to the
fact that after modding out the complex plane with $T$ the resulting
fundamental domain has infinite area (measured with respect to
$\frac{i}{2}\frac{d\tau\wedge d\bar\tau}{(\text{Im}\,\tau)^2}$), and this
leads to an infinite mass per volume element, see \eqref{total-mass}.  Thus,
to obtain solutions of finite mass, we are forced to include objects with
other monodromies.  Here we employ the S-duality of the theory.

We will focus on solutions whose monodromy group is $PSL(2,\mathbb{Z})$ which
is generated by $T$ and $S$. To show that one can also work with subgroups we
indicate in subsection \ref{subgroup} the construction of such solutions for
the specific case of the group $\Gamma_{0}(2)$ whose generators are $T$ and
$ST^2S$.  Having chosen a monodromy group we can specify the functions
$\tau(z)$ and $f(z)$.


\subsection{The function $\tau(z)$}\label{sec:tau}

Since $\tau$ and $\Lambda\tau$ are identified we need a function, $j(\tau)$,
which is monodromy neutral, i.e.~is an automorphic function of the monodromy
group

\begin{equation}
j(\Lambda\tau)=j(\tau)\,,
\end{equation}

\noindent where $\Lambda$ is any element of that  group.
The local expansions of the function $j$ around the fixed points of $\Lambda$
are as given in \eqref{taudetq=0} and \eqref{taudetq>0}.

\begin{figure}[htbp]
  \centering
  \includegraphics[scale=.5]{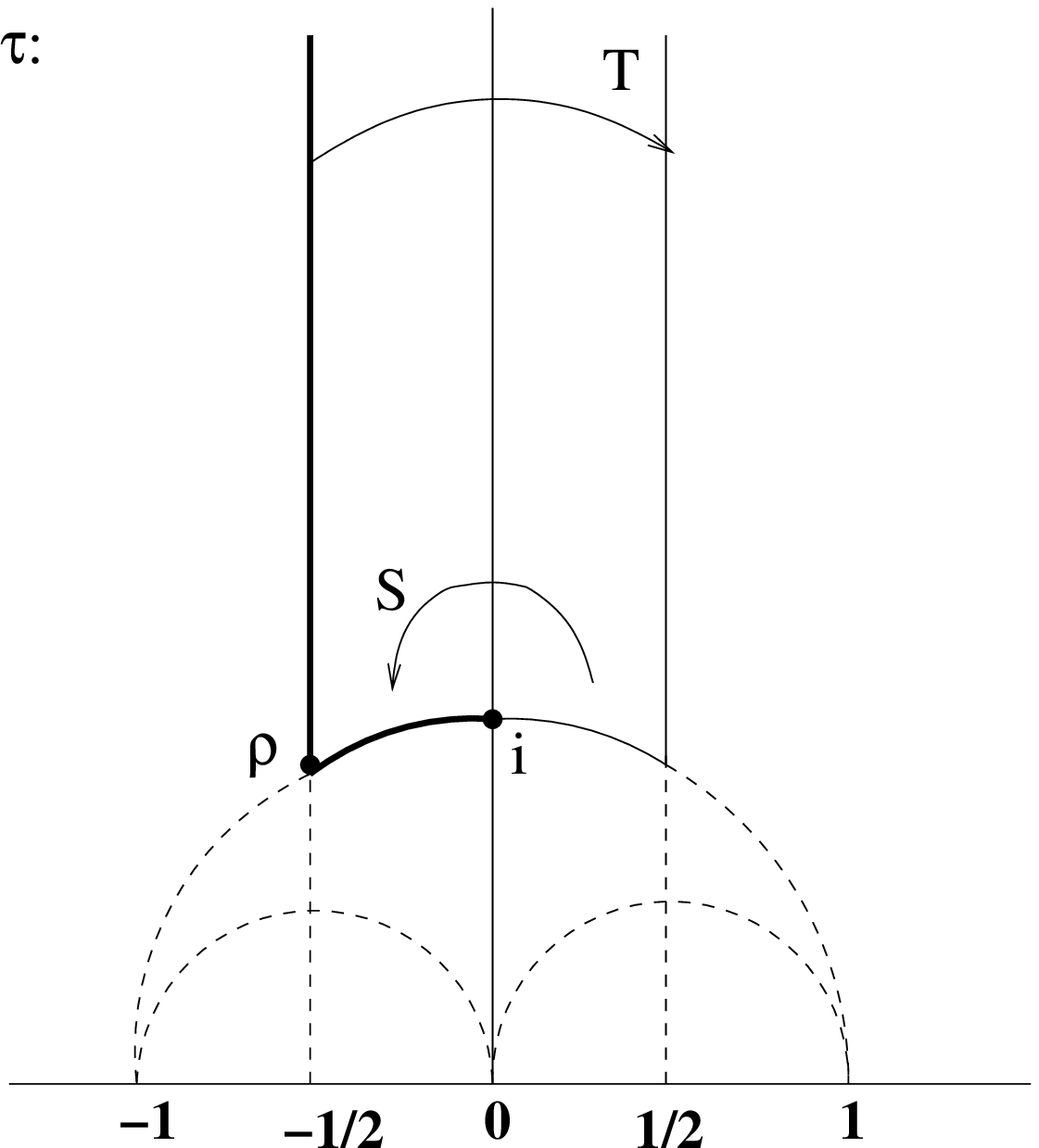}
\hspace{1cm}
  \includegraphics[scale=.5]{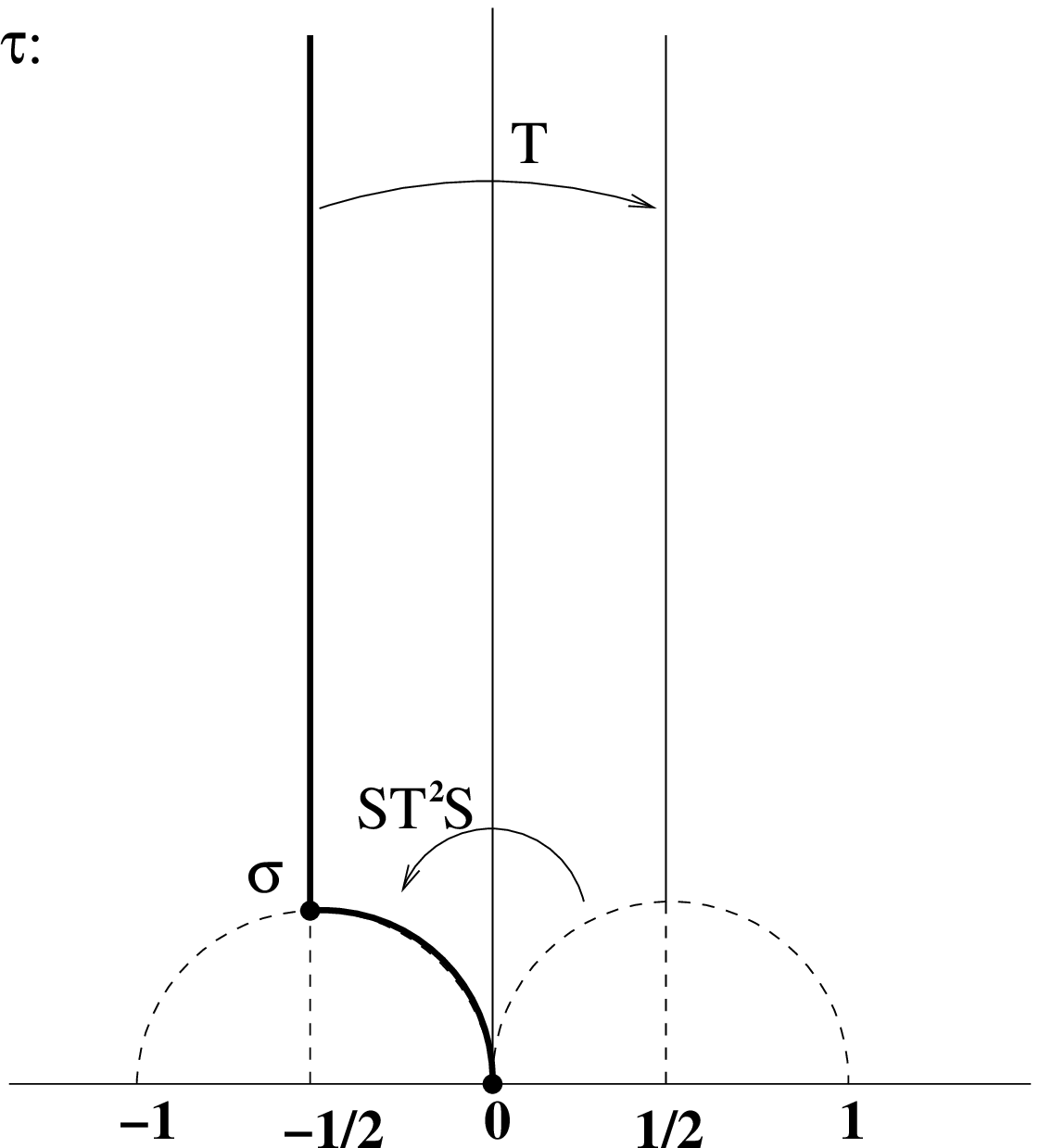}
 \caption{\it The fundamental domains of the groups $PSL(2,\mathbb{Z})$ and
   $\Gamma_{0}(2)$, respectively. The points $\rho$ and $\sigma$ denote the points
$-\frac{1}{2}+\frac{i}{2}\sqrt{3}$ and $-\frac{1}{2}+\frac{i}{2}$,
respectively. }
 \label{fig:funddom}
\end{figure}

A region of the complex upper half plain containing values of $\tau$ which are
inequivalent under the monodromy group but which are related to all the points
in the upper half plane is a fundamental domain of the monodromy group.  Note
that the fundamental domain is in general an orbifold.  Points which are fixed
points under some elements of the group are called orbifold points. In our
examples we always deal with a total of three orbifold points. In figure
\ref{fig:funddom} and table \ref{groupproperties} we have summarized some
properties of the monodromy groups $PSL(2,\mathbb{Z})$ and $\Gamma_0(2)$ and
the standard choices for their fundamental domains (see, e.g.~\cite{Serre}.)

We require that $j(\tau)$ maps the fundamental domain onto the Riemann sphere
$\hat{\mathbb{C}}$ in a one-to-one fashion, so that the inverse function
$j^{-1}$ exists.  The function $\tau(z)$ is then given by $\tau(z)=
j^{-1}(z)$.  Often we will include a further map from the Riemann sphere to
$N$ copies of itself which is given by the $N$ to 1 automorphism $z
\rightarrow P(z)/Q(z)$ for polynomials $P(z)$ and $Q(z)$. For $N=1$ these
polynomials are fixed by the requirement that the three orbifold points of the
fundamental domain are mapped to three given points in the $z$--plane, which
can always be achieved by an $SL(2,\mathbb{C})$ transformation. For instance,
the modular $j$ function maps the points $\{i\infty,\, \rho,\, i\}$ to
$\{\infty,\, 0,\, 1\}$ with $\rho=-\tfrac{1}{2}+\frac{i}{2}\sqrt{3}$.
Similarly the function $j_{\Gamma_{0}(2)}$ maps the points $\{i\infty,\,
\sigma,\, 0\}$ to $\{\infty,\, 0,\, 1\}$ with
$\sigma=-\tfrac{1}{2}+\frac{i}{2}$.  For $N>1$ the polynomials $P(z)$ and
$Q(z)$ are fixed by the further requirement of how many branes are placed
at the three points $z_{i\infty}, z_\rho$ and $z_i$ where the subscript
indicates the value of $\tau$ at that point. For $N=1$ there is one brane at
each point. In the next section we will give explicit realizations of $P(z)$
and $Q(z)$. Note that, for general $N$, the mass formula \eqref{energy}
becomes

\begin{equation}
m=\frac{1}{16\pi G^{(3)}_{N}}\left(N\times\text{area fundamental
domain}+ \ 2\sum_{j}\,\delta_j
    \right)\, ,
  \label{total-mass}
\end{equation}

\noindent where the area is measured with the area element

\begin{equation}
{\textstyle\frac{i}{2}}\frac{d\tau\wedge
d\bar\tau}{(\text{Im}\,\tau)^2}\, .
\end{equation}

Summarizing, we have the sequence of maps

\begin{equation}
z\hskip 1truecm \stackrel{N\rightarrow 1}{\longrightarrow} \hskip
1truecm \frac{P(z)}{Q(z)}\hskip 1truecm
\stackrel{j^{-1}}{\longrightarrow}\hskip 1truecm \tau (z) =
j^{-1}\left(\frac{P(z)}{Q(z)}\right)\,.
\end{equation}

The inverse mapping $j^{-1}$ which maps from the Riemann sphere
$\hat{\mathbb{C}}$ onto the fundamental domain has branch cuts connecting the
points $z_{i\infty}$ to $z_{\rho}$ and $z_{\rho}$ to $z_i$. Likewise the
inverse function $j_{\Gamma_{0}(2)}^{-1}$ has branch cuts connecting the
points $z_{i\infty}$ to $z_{\sigma}$ and $z_{\sigma}$ to $z_{0}$.



\vskip .3truecm

\begin{table}[h]
\hspace{-.5cm}
\begin{tabular}{||c||c|c|c||c|c||}
  \hline
   & {\small generators} & {\small orbifold pts.} &  {\small area}  & $j(\tau)$ & $F(\tau)$ \\
  \hline \hline
    &&&&&\\
  $PSL(2,\mathbb{Z})$ & $T, S$ & $i\infty,\, \rho,\, i$ & $\pi/3$ & Klein's modular $j$ function \cite{Apostol} & $\eta^2(\tau)$  \\
                            &&&&&\\
  \hline
    &&&&&\\
  $\Gamma_{0}(2)$ & $T, ST^2S$ & $i\infty,\, 0,\, \sigma$ & $\pi$ &
${\displaystyle
j_{\Gamma_{0}(2)}\equiv\frac{1}{(1+i)^{12}}\left(\frac{\eta(\tau)}{\eta(2\tau)}\right)^{24}}$
&
  $\eta(\tau)\eta(2\tau)$ \\
  &&&&&\\
  \hline
\end{tabular}
\caption{Properties of the groups $PSL(2,\mathbb{Z})$ and
$\Gamma_{0}(2)$ and corresponding realizations of the functions
$j(\tau)$ and $F(\tau)$.} \label{groupproperties}
\end{table}


\subsection{The function $f(z)$}\label{f}

The function $f(z)$ can be written in the form

\begin{equation}
\label{generalformf} f(z)=F(\tau)h(z)\, ,
\end{equation}

\noindent where $F(\tau)$ is a modular function which transforms
under $PSL(2,\mathbb{Z})$ as

\begin{equation}
\label{trafoF}
F\left(\frac{a\tau+b}{c\tau+d}\right)=e^{i\beta(a,b,c,d,\tau)}
(c\tau+d)F(\tau)\, ,
\end{equation}

\noindent and $h(z)$ is a function of z which we choose such that
when going around a 7-brane it transforms as

\begin{equation}\label{monh}
h(z)\rightarrow e^{i k\pi} e^{-i\beta(a,b,c,d,\tau(z))} h(z)
\end{equation}

\noindent where $k=0,1$. For $k=0$ we have a plus sign and this
must be used for transformations under the identity of
$SL(2,\mathbb{Z})$. For $k=1$ we have a minus sign which must be
used for transformation under the element $-\mathbbm{1}$ of
$SL(2,\mathbb{Z})$. This additional sign is necessary because $f$
transforms under $SL(2,\mathbb{Z})$. It will play an important
role later in the construction of solutions. We see that $f$
transforms under $SL(2,\mathbb{Z})$ as $\pm\mathbbm{1}\times
PSL(2,\mathbb{Z})$ and that the $\pm\mathbbm{1}$ part is
independent of $\tau$.

When going around a D7-brane the function $f$ does not change, so
that $f$ has no zero/pole at the point $z=z_{i\infty}$. This
follows from equation \eqref{holonomy}. We do however expect to
measure a positive contribution to the total energy when going to
infinity, $\vert z\vert\rightarrow\infty$, due to the presence of
a D7-brane. Therefore, $f$ must lead to a non-trivial deficit
angle at infinity.  Indeed, we know that $f(z) \rightarrow z^{-4
G_{N}^{(3)}m}$ for $\vert z\vert \rightarrow \infty$. This
asymptotic behaviour must follow from the function $h(z)$ since
$\tau_{\infty}$ = cst and therefore $F(\tau)$ does not contribute.
Around $z=z_{i\infty}$ this leads to a factor
$(z-z_{i\infty})^{-\alpha}$ with $\alpha>0$ in $h(z)$, and
therefore to a pole in $h(z)$ for $z=z_{i\infty}$. Since we just
argued that $f(z)$ cannot have such a pole, it must be cancelled
by a zero of $F$ at $\tau=i\infty$. We have thus established that
$F(\tau)$ must be a cusp form. A lot is known about such cusp
forms in the mathematical literature. The explicit realizations of
these cusp forms in terms of the Dedekind eta function
$\eta(\tau)$ for the groups $PSL(2,\mathbb{Z})$ and $\Gamma_0(2)$
is given in table \ref{groupproperties}. Using the monodromies of
this cusp form and the required monodromies of $f(z)$ it is not
difficult to derive an explicit realization of the function
$h(z)$.

The choice of the function $h(z)$ is case dependent. We will give explicit
expressions in section \ref{sec:examples}. Here we only give the
transformation of the Dedekind $\eta$-function under the different
$PSL(2,\mathbb{Z})$-transformations:

\begin{alignat}{2}
T: \quad & \eta^2(\tau+1) &&
=e^{\pi i / 6}\ \eta^2(\tau)\,,\label{Ttr} \\
& && \nonumber \\
S: \quad & \eta^2\left(-\frac{1}{\tau}\right) && = e^{- \pi i /2}\
\tau\ \eta^2(\tau)\,,
\label{Str} \\
& && \nonumber \\
T^{-1}S: \quad & \eta^2\left(-\frac{\tau+1}{\tau}\right) && =e^{-
2\pi i /3}\ \tau\ \eta^2(\tau)\,.\label{STtr}
\end{alignat}

For the convenience of the reader we present in table
\ref{anglesofKillingspinor} the monodromies of $\tau$ and $f$ measured when
going around the points $z_{i\infty},\,z_{\rho},\,z_i$ in a counter clockwise
direction and the deficit angles. Instead of comparing monodromies it is
sometimes convenient to compare deficit angles.

\begin{table}[h]
\begin{center}
\begin{tabular}{||c||c|c|c|c||}
  \hline
  location  & $SL(2,\mathbb{Z})$ & $(p,q,r)$ & monodromy $f$ &deficit angle $\delta$\\
  \hline \hline
  &&&&\\
  $z_{i\infty}$ & $T$ & $(1,0,0)$  & $f\rightarrow f$     &0 \\
  &&&&\\
  \hline
  &&&&\\
  $z_i$ & $S$ & $(-\pi/2,-\pi/2,0)$ &  $f\rightarrow \tau f$ &
-$\pi/2$ \\
  &&&&\\
  \hline
  &&&&\\
  $z_i$  & $-S$ & $(\pi/2,\pi/2,0)$ &  $f\rightarrow e^{i\pi}\tau f$&
 $\pi/2$\\
  &&&&\\
  \hline
  &&&&\\
  $z_{\rho}$  & $T^{-1}S$ & $(-\frac{4\pi}{3\sqrt{3}},-\frac{4\pi}{3\sqrt{3}},-\frac{4\pi}{3\sqrt{3}})$
  & $f\rightarrow \tau f$ &-$2\pi/3$ \\
  &&&&\\
  \hline
  &&&&\\
  $z_{\rho}$  & $-T^{-1}S$ & $(\frac{2\pi}{3\sqrt{3}},\frac{2\pi}{3\sqrt{3}},\frac{2\pi}{3\sqrt{3}})$
    & $f\rightarrow e^{i\pi}\tau f$  &$\pi/3$ \\
  &&&&\\
  \hline
\end{tabular}
\caption{The monodromy of $\tau$ and $f$,  the $p,q,r$ values and
the deficit angles for $\tau=i\infty,\,\rho,\,i$. The deficit
angle $\delta$ is computed using
$\delta=\text{sign}(q)\,\sqrt{\text{det}\,Q}$. In the monodromy
transformation for $f$ we take $1=e^{i0}$ and $-1=e^{i\pi}$ in
agreement with the discussion below equation \eqref{monh} about
the number $k=0,1$. }\label{anglesofKillingspinor}
\end{center}
\end{table}

This completes the construction of a globally well--defined
7-brane solution. We will present several explicit examples in the
next section.

\section{Examples of Solutions}\label{sec:examples} %

In this section we give several examples of globally
supersymmetric solutions using the ingredients discussed so far.
In particular, we will analyze the global properties of the
Killing spinor. This analysis will determine the precise form of
the function $h(z)$.

We will concern ourselves mostly with the monodromy group
$SL(2,\mathbb{Z})$. Only in the last subsection \ref{subgroup} an
example with the monodromy group $\Gamma_{0}(2)$ will be
discussed.


\subsection{Solutions containing a single D7-brane}

We first present the simplest possible solution containing a
single D7-brane using the results of the previous sections. In the
next subsection we will argue that the solutions of this
subsection can be viewed as special limits of more general
solutions. The latter solutions which we refer to as the basic
building blocks can be used to generate all possible 7-brane
solutions containing an arbitrary number of D7-branes.

One way to derive an explicit form for $h(z)$ is to compare
monodromies.  We first consider a D7-brane which is located at the
point $z_{i \infty}$. The monodromy of $f=\eta^2$ around the point
$z_{i\infty}$ is $f \rightarrow e^{\pi i/6}f$, according to
eq.~\eqref{Ttr}. This does not coincide with the transformation
required by $SL(2,\mathbb{Z})$, which is $f \rightarrow f$.  For
this reason one should include a factor $(z -
z_{i\infty})^{-1/12}$ in $f$, i.e. $h(z) \sim (z -
z_{i\infty})^{-1/12}$. With this choice of $f$ the behaviour of
the Killing spinor around the point $z_{i\infty}$ coincides with
the $SL(2,\mathbb{Z})$ requirement.

Branes located at the points $z_i$ and $z_{\rho}$ are named after
their monodromy under $SL(2,\mathbb{Z})$. For example, if we
consider a point $z_i$ with $S$ monodromy then we call this an
$S$-brane. For such a brane we must have $f\rightarrow \tau f$.
From equation \eqref{Str} it follows that around $z_i$ the
function $h$ must transform as $h\rightarrow e^{i\pi/2}h$, so we
include an additional factor $(z-z_i)^{1/4}$. For a $(-S)$-brane
one must include a factor of $(z-z_i)^{-1/4}$. If we consider a
brane at $z_{\rho}$ with $T^{-1}S$ monodromy then
$f\rightarrow\tau f$. In order to compensate the factor which
appears in \eqref{STtr} we must include a factor
$(z-z_{\rho})^{1/3}$. Likewise for a brane with $-T^{-1}S$
monodromy the factor which appears in $h$ is
$(z-z_{\rho})^{-1/6}$.

If we take $z_{\rho}=\infty$, which means that the asymptotic
value of $\tau$ is $\rho$, the simplest solution is given by

\begin{align}\label{N=1solution-zrho}
j = \frac{(z_i-z_{i \infty})}{(z - z_{i \infty})} \,, \quad  f =
\eta^2 (z - z_{i \infty})^{-1/12} (z - z_i)^{-1/4} \,.
\end{align}

\noindent This solution is asymptotically conical with a deficit
angle of $2 \pi / 3$. If on the other hand $z_i=\infty$, so that
the asymptotic value of $\tau$ is equal to $i$ then we find

\begin{align}\label{N=1solution-zi}
j = \frac{(z-z_\rho)}{(z - z_{i \infty})} \,, \quad  f
= \eta^2 (z - z_{i \infty})^{-1/12} (z - z_\rho)^{-1/6}  \,. %
\end{align}

\noindent This is asymptotically a cone with deficit angle $\pi /
2$. The properties of solutions of this type will be discussed
further at the end of the next subsection.


\subsection{The basic building blocks}\label{blocks}

In order to consider more general solutions for $N>1$ it would be
convenient to have a means of obtaining them directly starting
from the $N=1$ solutions. The above method of constructing
solutions via monodromy requirements becomes cumbersome for large
$N$. In this subsection we will construct two basic solutions, the
$N=1A$ and $N=1B$ solutions also called the basic building blocks,
out of which any other solution can be generated in a manner to be
described shortly. In particular the solutions presented in the
previous subsection appear as special limits of these $N=1A$ and
$N=1B$ solutions.

The problem in using \eqref{N=1solution-zrho} and
\eqref{N=1solution-zi} as the starting point for higher $N$
solutions is that they have a nontrivial asymptotic geometry. We
can however add a point mass to a solution whose mass is equal and
opposite to the total mass measured at infinity. This is because
point masses in 2+1 dimensions only form deficit angles, but
otherwise do not deform the solution. The resulting space is then
by construction asymptotically flat and has total mass zero. In
terms of monodromies the idea is thus to start with a solution in
which all points around which there is a nontrivial monodromy are
at finite values of $z$. So there is no monodromy for $\tau$ and
$f$ around the point $z=\infty$.

For $N=1$ two such configurations are possible\footnote{We do not
consider the possibility of having branes with $-T$ monodromy
because $-T$ is not continuously related to the identity in
$SL(2,\mathbb{Z})$.}. The first consists of three branes with
monodromies $T$, $S$ and $- T^{-1} S$ and the following choice for
the functions $j$ and $f$:

 \begin{align}
 1A:\hskip 1truecm
 j = \frac{(z - z_\rho)(z_i - z_{i \infty})}{(z - z_{i \infty})(z_i - z_\rho)} \,,
 \quad  f = \eta^2 (z - z_{i \infty})^{-1/12} (z - z_i)^{1/4} (z - z_\rho)^{-1/6}
 \,.\label{N=1A}
 \end{align}

\noindent Due to the zeroes and poles of $f$, there are deficit
angles (and hence masses) at the points $z_i$ and $z_\rho$ which
are given by $- \pi /2$ and $\pi / 3$, respectively. Note that the
brane with $S$ monodromy has a negative mass and deficit angle,
and that the total mass adds up to zero:

\begin{equation}
m \sim \frac{\pi}{3} + \frac{2\pi}{3} - \pi = 0\,.
\end{equation}

\noindent We will refer to this solution as the $N=1A$ solution.
In the limit in which the negative mass brane is sent to the
asymptotic region\footnote{To take this limit one must rescale $f$
in \eqref{N=1A} with a
  factor $z_i{}^{-1/4}$ in order that this $f$ goes to the $f$ of
  \eqref{N=1solution-zi}.\label{limitingprocedure}}  $z_i\rightarrow \infty$ the solution becomes
equation \eqref{N=1solution-zi}. Whenever one constructs solutions
in this way such a limiting procedure must always be performed.
The procedure should be considered as a solution generating
technique. It does not represent some kind of physical process.

The method used here is somewhat similar to the method of images
used in section \ref{sec:energy}. The difference is that here we
are considering solutions that are asymptotically $\mathbb{R}^2$
whereas the image object gives rise to $S^2$. Using the latter
method the 1A solution would have contained the factor $(z -
z_i)^{-2+1/4}$ instead of $(z - z_i)^{1/4}$ leading to a total
deficit angle $4\pi$.  In this case the point $z_i$ satisfies the
criterion given in footnote \ref{asymptoticregion} of section
\ref{sec:energy} and should be considered as the asymptotic
region. In the present case the point $z_i$ does not satisfy this
criterion and it should be considered as the location of an actual
object with negative deficit angle equal to $\delta_{\infty}$ (see
equation \eqref{BPSbound}). In order to do away with it one has to
take a singular limit as described in footnote
\ref{limitingprocedure}.

\begin{figure}[htbp]
  \centering
\includegraphics[scale=.5]{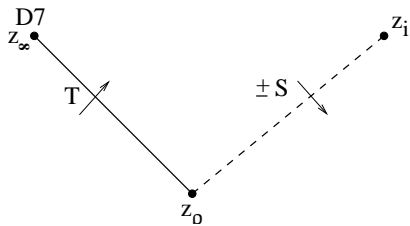}
 \caption{\it The $N=1A$ and $N=1B$ supersymmetric 7-brane solutions,
   with the D7-brane at $z_{i \infty}$ and $\text{det}\,Q>0$-branes at the
   points $z_\rho$ and $z_i$. The filled (dashed) lines are $T$ ($S$) branch
   cuts. The sign choice corresponds to taking $k=0$ $(\mathbbm{1})$ and $k=1$
   $(-\mathbbm{1})$ in \eqref{monh}.  The $N=1A$ and $N=1B$ solutions have
   upper and lower signs, respectively. }
 \label{fig:N=1}
\end{figure}

The second possibility consists of branes with monodromies $T$,
$-S$ and $ T^{-1} S$. Note that these only differ in a number of
signs from the previous one. Since the scalars are only sensitive
to the $PSL(2,\mathbb{Z})$ part the function $j$ remains
unchanged. The Killing spinor is sensitive to the signs and
therefore the function $f$ is different in this case. It is given
by

\begin{align}
1B:\hskip 2truecm  f = \eta^2 (z - z_{i \infty})^{-1/12} (z -
z_i)^{-1/4} (z - z_\rho)^{1/3} \,.
\end{align}

\noindent The poles and zeroes of $f$ now give rise to deficit
angles of $\pi /2$ and $- 2 \pi / 3$ at the points $z_i$ and $z_{\rho}$,
respectively. Note that in this case the object at $z_\rho$ has a negative
mass and again the total mass is vanishing:

\begin{equation}
m \sim \frac{\pi}{3} - \frac{4\pi}{3} +\pi = 0\,.
\end{equation}

\noindent This solution will be referred to as the $N=1B$
solution. The limit\footnote{In this case one must first multiply $f$ by a
  factor $z_\rho{}^{-1/3}$ before taking the limit.} $z_\rho \rightarrow
\infty$ gives rise to equation \eqref{N=1solution-zrho}. The 1A and 1B
solutions are pictorially represented as in figure \ref{fig:N=1}.

One should think of figure \ref{fig:N=1} as being the Riemann
sphere parameterized by $z$. When crossing a branch cut we go to a
different sheet on the Riemann sphere. Lines of $T$ monodromy
connect one point $z_{i\infty}$ to one point $z_{\rho}$ and lines
of $S$ monodromy connect one point $z_{\rho}$ to one point $z_i$.
Just from the order of the branch cuts one can uniquely construct
the function $j(\tau)$ from the figure. In order to also construct
the function $f$ we have indicated the appropriate signs which
play a role in determining the function $h$ as explained below
equation \eqref{monh}. In other words the figure is a unique
pictorial representation of the pair $(\tau,f)$. A particularly
convenient property of such a representation is that it also
captures the global positioning of the branch cuts and thus the
monodromy measured when going along large loops. This should be
contrasted with equation \eqref{localmonodromytau} where the
monodromy of $\tau$ is computed for loops which go around a brane
at an infinitesimal distance.

Solutions with higher $N$ can now be constructed by combining any
given number of the 1A solutions with any given number of the 1B
solutions. For example, for $N=2$ one can combine two $N=1$
solutions of the same or different types, yielding three different
possibilities. The combination of e.g.~two $1A$ solutions leads
to\footnote{The combination of one $1A$ and one
  $1B$ solution is given in \eqref{N=2solution}.}

 \begin{align}
  & j = \lambda \, \Pi_{n=1,2} \frac{ (z-z^{(n)}_\rho)}{(z - z^{(n)}_{i \infty})} \,, \quad
  & f = \eta^2 \, \Pi_{n=1,2} (z - z^{(n)}_{i \infty})^{-1/12} (z - z_i^{(n)})^{1/4} (z -
  z_\rho^{(n)})^{-1/6} \,,\label{jfN=2solution}
 \end{align}

\noindent where $\lambda$ is some complex constant. The two points
$z^{(n)}_i$ are given by the roots of the equation $j=1$ and depend on the
value of $\lambda$. The other two possibilities with $N=2$ require the obvious
changes in the powers of $(z - z_{i})$ and $(z-z_{\rho})$ in the function $f$.
Higher-$N$ solutions are built along the same lines. Equation
\eqref{jfN=2solution} shows the advantage of working in a space of total
deficit angle zero as opposed to one of total deficit angle of $4\pi$ because
only in the former case can we write the function $h$ which appears in $f$ as
a product of $h(1A)$ and $h(1B)$.

Several remarks are in order. In the 7-brane solutions of
\cite{Greene:1989ya,Gibbons:1995vg} the largest number of
D7-branes which a solution can contain is 24 because for this
number the transverse space has become a sphere. It is now
possible to construct solutions with more than 24 D7-branes if and
only if one allows for the presence of $\text{det}\,Q>0$-branes
with negative mass. Further, in order to have the value of $\tau$
at infinity arbitrary so that the asymptotic region can be taken
as an approximation of perturbative string theory we need to take
combinations of the 1A and 1B solutions such that there is one
point whose $\tau$ monodromy is the identity (in
$PSL(2,\mathbb{Z}$)). This can for example be realized by taking
two points with $S$ monodromy coincident and subsequently sending
that point to infinity. The third remark is that because $T$
monodromies are of infinite order one cannot eliminate the
D7-branes. The reason that the $\text{det}\,Q>0$-branes can be
eliminated is because their monodromies are of finite order.

Note that both $N=1$ solutions are characterized by three complex
constants, while the $N=2$ solution has five. A general solution
containing $N$ D7-branes has $2N +1$ complex parameters. Equation
\eqref{globalcoordinatefreedom} tells us that the metric is fixed
up to global $SL(2,\mathbb{C})$ transformations of the complex
coordinate $z$. By placing the branes all at finite values of $z$
we have fixed the $z\rightarrow 1/(z-c)$ element of
$SL(2,\mathbb{C})$. By using the freedom to shift and scale $z$ we
can eliminate two complex constants leaving $2N-1$ free. The
number of complex parameters $2N-1$ is further reduced by one if
we send the point with negative deficit angle $\delta_{\infty}$
off to infinity. This fixes the $SL(2,\mathbb{C})$ coordinate
freedom. From the Killing spinor equation \eqref{Killingspinor} we
see that we have the additional freedom to scale the absolute
value of the function $f$. This would leave the metric invariant
if we could compensate by a scaling of $z$, which cannot be done
since we have already used up the $SL(2,\mathbb{C})$ coordinate
freedom. We leave, then, this modulus free, and we end up with
$2(2N-2)+1$ free real parameters. This number is made out of
$2N-2$ complex parameters and one real parameter. The complex
numbers describe the relative positions of the branes and will at
the end of this subsection be related to the number of vectors
which can exist on these 7-brane configurations.

The background fields of type IIB which may have nontrivial zero
modes on the 7-brane backgrounds are $g_{\mu\nu}$,
$A_{\mu\nu\rho\sigma}^{+}$ (self-dual 4-form) and $\tau$. The
2-forms do not appear since in order to construct the solutions we
had to divide out type IIB supergravity by $SL(2,\mathbb{Z})$. The
fact that $\tau$ is a very special function of $z$, which it must
be in order that it takes values in the fundamental domain of
$PSL(2,\mathbb{Z})$, suggests to interpret the excitations of the
background as the modes of a supergravity theory in 8 dimensions.
It was mentioned in footnote \ref{dividingbySL2Z}, section
\ref{sec:holonomy}, that this theory has $N=1$ supersymmetry and
consists of one supergravity and one vector multiplet.  The above
counting argument seems to suggest that this 8-dimensional
supergravity must be coupled to an additional number of $2N-2$
vector multiplets coming from the number of free complex numbers
describing the relative positions of the branes. Notice that for
$N\ge 3$ this number is larger than the number of D7-branes, $N$,
and hence, there are more vectors than D7-branes. This seems to
suggest that one has to attribute some of the vectors to the
worldvolume theories of the $\det\, Q>0$-branes.


\subsection{F-theory solutions} \label{D7-trunc}

The well--known 7-brane configurations of F-theory have the
property that the monodromy of $\tau$ close to the points $z_i$,
$z_{\rho}$ is the identity in $PSL(2,\mathbb{Z})$ and $T$ around
$z_{i\infty}$. Further it is required that the function $f$ has no
zeros, which can be interpreted as saying that all mass comes
solely from the D7-branes. This condition is satisfied if and only
if $\tau$ and $f$ are of the following form

\begin{equation}\label{Ftheorytau}
j(\tau) =\frac{P^3(z)}{P^3(z)+Q^2(z)} \,, \quad
f=\eta^2\left(P^3+Q^2\right)^{-1/12}\,,
\end{equation}

\noindent where $P^3+Q^2$ is a polynomial of order $N$ whose zeros
are the locations of the D7-branes. Solutions of this type exist
whenever $N$ can be divided by either 2 or 3. Thus they exist for
$N=2,3,4,6,8,9,10,12,24$. In F-theory one cannot go beyond $N=24$
because for this value of $N$ the transverse space has become a
sphere. Each time the deficit angle at infinity is given by $2\pi
N/12$. Out of this set of solutions of $\tau$ those which have
$N=6,12,24$ are such that the asymptotic value of $\tau$ can take
any value. For the other solutions the asymptotic value is
necessarily either $i$ or $\rho$. Therefore, these are necessarily
nonperturbative in nature. Functions $\tau$ which solve equation
\eqref{Ftheorytau} are the modular parameter of an elliptically
fibered torus whose base manifold is the transverse space of the
D7-branes. The geometric interpretation of $f$ is that $fdzd\tau$
becomes the holomorphic $(2,0)$-form of the CY two--fold.

F-theory solutions form a subset of our general solutions.  They
are obtained by having coincident branes of the same
type\footnote{It is not possible to
  have branes of different type coinciding, since these require different
  values of $\tau$ at the same point.} as follows. Take combinations of the 1A and
1B solutions in which there is only one point $z_i$ or $z_{\rho}$ which has a
negative deficit angle. Around this special point the monodromy of $\tau$ can
be anything. Further, there should be no points $z_i$ and $z_{\rho}$ around
which either $\tau$ or $f$ has a nontrivial monodromy.
\newline

\noindent \textbf{The $N=2,3$ solution with only D7-branes}
\newline

Branes at the points $z_i$ either have deficit angle $- \pi / 2$ or $+ \pi /
2$. Combining two branes one of monodromy $S$ and one of monodromy $S^{-1}=-S$
gives rise to a cancellation of their masses. If we combine the $N=1A$ and
$N=1B$ solutions without making any assumption about the positions of the
branes we have for the function $f$

\begin{equation}
f = \eta^2 \, (z - z^{(1)}_{i \infty})^{-1/12}
(z - z^{(2)}_{i \infty})^{-1/12} (z - z_i^{(1)})^{1/4}
(z - z_i^{(2)})^{-1/4}(z - z_\rho^{(1)})^{-1/6}(z - z_\rho^{(2)})^{1/3} \,.
\label{N=2solution}
\end{equation}

The cancellation mechanism just discussed can now be applied by taking the
points $z_i^{(1)}$ and $z_i^{(2)}$ coincident. Next we also take the points
$z_\rho^{(1)}$ and $z_\rho^{(2)}$ coinciding so that we end up with only one
point $z_{\rho}$ around which $\tau$ has a nontrivial monodromy. Finally, one
must multiply the resulting function $f$ by $z_{\rho}^{-1/6}$ and send
$z_{\rho}\rightarrow\infty$. This leaves us with the required form for $f$

\begin{equation}
  f = \eta^2 \, (z - z^{(1)}_{i \infty})^{-1/12} (z - z^{(2)}_{i
  \infty})^{-1/12}\,.
 \end{equation}

\noindent Applying these choices to the function $j(\tau)$ for a general
$N=2$ solution which is given in \eqref{N=2solution} leads to

\begin{equation}
    j(\tau)=\frac{C}{C+(z-z_{i})^2}\,,
\end{equation}

\noindent where $C\neq 0$ is a constant. It is clear that the asymptotic
value for $\tau$ is equal to $\rho$. This solution is therefore strictly
nonperturbative.

In a similar fashion one can construct the $N=3$ solution. Now one
can combine two branes with monodromy $-T^{-1}S$ and one brane of
monodromy $T^{-1}S$ at $z_\rho$ leading to vanishing mass and
trivial monodromy. This is based on the deficit angles $+ \pi / 3$
or $- 2 \pi / 3$ and the identity $(- T^{-1} S) \cdot (-T^{-1} S)
\cdot (T^{-1} S) = \mathbbm{1}$. Hence in this case we need to
take two $N=1A$ and one $N=1B$ solutions. Doing so we have three
points $z_i$ with monodromies $S$, $S$ and $-S$ that we need to
deal with. Taking one brane with $S$ and one brane with $-S$
monodromy coincident we are left with one point $z_i$ which has
$S$ monodromy and hence a negative deficit angle. Sending this
point to infinity we end up with the form of $f$ and $\tau$ as in
\eqref{Ftheorytau}. This time the asymptotic value for $\tau$ is
equal to $i$.\newline

\noindent \textbf{The $N=6$ solution with only D7-branes}
\newline

The case $N=6$ is the first instance in which solutions with only
D7-branes and an arbitrary value for $\tau$ at infinity are
possible.  It can be constructed as follows. From four $N=1A$
solutions and two $N=1B$ solutions one can make the following
combinations of points:

\begin{align}
  N=6: \quad
  \begin{cases}
   \begin{array}{cc}
   (- T^{-1} S) \cdot (-T^{-1} S) \cdot (T^{-1} S) = \mathbbm{1} \,, \quad &
\text{at two points~}z_\rho \,, \\
   (- S) \cdot S = \mathbbm{1} \,, \quad & \text{at two points~}z_i \,, \\
    S \cdot S  = - \mathbbm{1} \,, \quad & \text{at one point~}z_i \,.
   \end{array}
  \end{cases}
 \label{coinciding}
\end{align}

\noindent The function $f$ corresponding to \eqref{coinciding} is
given by

\begin{equation}\label{fN=6}
    f=\eta^2(P^3+Q^2)^{-1/12}(z-z_i)^{1/2}\,.
\end{equation}

\begin{figure}[htbp]
  \centering
\includegraphics[scale=.5]{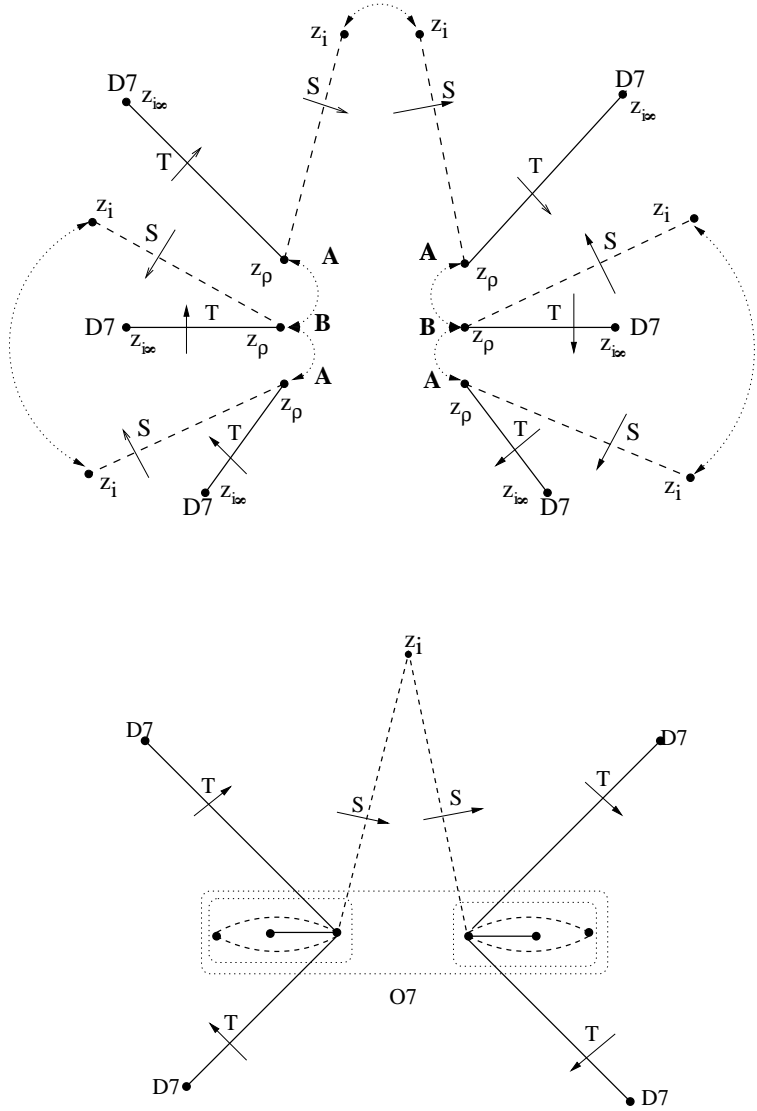}
 \caption{\it The most general supersymmetric $N=6$ solution (lower diagram)
   with only non-trivial $T$-monodromies around the points $z_{i \infty}$,
   where the D7-branes are located can be constructed from six elementary
   $N=1$ solutions (upper diagram) by forcing some branes to coincide. The
   filled (dashed) lines are $T$ ($S$) branch cuts while the dotted lines in
   the upper diagram relate the points which are made coincident to obtain the
   lower diagram.}
 \label{fig:N=6}
\end{figure}

\noindent Note that four of the five  points have total
$SL(2,\mathbb{Z})$-monodromy $\mathbbm{1}$: the mass and charge
cancel here. The remaining point $z_i$ has $-\mathbbm{1}$ and
there is a negative mass at this point. See figure \ref{fig:N=6}
for an illustration. To obtain a solution with only D7-branes we
first multiply $f$ in \eqref{fN=6} by a factor $z_i{}^{-1/2}$ and
then send the point $z_i$ in $f$ to infinity. It should be
mentioned that the value of $\tau$ in the limit
$z_i\rightarrow\infty$ can now be different from $\tau=i$. The
value of $\tau$ after taking the limit depends on the values of
the constants appearing in the polynomials $P$ and $Q$.

There is a similar solution composed of three $N=1A$ and three
$N=1B$ solutions in which there are three $z_i$ and one $z_\rho$
orbifold point with vanishing mass and trivial monodromy $+
\mathbbm{1}$. The other point $z_\rho$ again has a monodromy $-
\mathbbm{1}$ and a negative mass. In fact this solution is
equivalent to the $N=6$ solution above: they differ in the
assignation of signs to the separate $N=1A$ and $N=1B$ solutions
but these all cancel in this particular configuration. In what
follows we will use the first parametrization of this solution.

A string theory interpretation of this solution has been proposed
in \cite{Sen:1996vd} (see also \cite{Dabholkar:1997zd}). It was
argued that the $N=6$ case can be seen as four D7-branes
accompanied by an orientifold O7-plane. The latter is obtained by
modding out $\text{Mink}_{1,7} \times \mathbb{R}^2$ with the
$\mathbb{Z}_2$-symmetry $(-)^{F_L} \Omega \,I_{8,9}  = - T^{-4}$
and hence it carries $-4$ units of D7-brane charge. The
combination $(-)^{F_L} \Omega$ acting on the world-sheet fields of
the type IIB superstring has the same effect as a $-\mathbbm{1}$
transformation has on the space--time fields of type IIB
supergravity. In the orientifold limit, the four D7-branes
coincide with the O7-plane and cancel the R-R charge. The
remaining monodromy is $- \mathbbm{1}$.

In the $N=6$ solution, the four separate branes with $T$-monodromy
in figure \ref{fig:N=6} are to be interpreted as the D7-branes. In
addition, there are two composite objects with monodromy $T^{-2}
S^{-1}$ and $S^{-1} T^{-2}$. These monodromies can be written as

\begin{align}
  M_{1,2} T M_{1,2}^{-1} \text{~~~ with ~~~}
  M_1 = \pm \left( \begin{array}{cc} 1 & \lambda_1 \\ -1 & 1-
  \lambda_1\end{array} \right)  \,, \quad
  M_2 = \pm \left( \begin{array}{cc} 1 & \lambda_2 \\ 1 & 1 +
  \lambda_2 \end{array} \right)  \,.
\end{align}

\noindent respectively. Due to the above relations these branes
can be viewed as $SL(2,\mathbb{Z})$-transformed versions of
D7-branes. Instead of $\tau \rightarrow i \infty$ the complex
scalar goes to the real line in their vicinity, and henceforth IIB
perturbative string theory is not valid there. Only the monodromy
at infinity has a perturbative meaning. For this reason the two
composite objects should be considered as one single object,
i.e.~the O7-plane. In the perturbative limit this composite object
indeed has a monodromy given by $- T^{-4}$. It is clear from the
picture that one can take at most four out of the six D7-branes
coincident.

The orientifold limit corresponds to taking the most general
solution of the equation $j(\tau)=\text{cst}$. For the $N=6$
solution this means that all points $z_{i, \rho, i \infty} \neq
\infty$ are coincident at some finite value for $z$. The only
remnant is the deficit angle of $\pi$ and the $SL(2,\mathbb{Z})$
transformation $- \mathbbm{1}$ around this point. There does not
exist a limit in which the O7-plane shrinks to a point while the
four D7-branes remain at finite distances from the orientifold
point.\newline

\noindent \textbf{The $N=12,24$ solutions with only D7-branes}
\newline

The $N=12,24$ solutions can be constructed in more than one way.
For example the $N=24$ solution may be constructed by gluing four
$N=6$ solutions. This leads to a configuration which at infinity
(which in this case is a perturbative region) seems to be composed
out of four groups of O7-planes each of which is accompanied by
four D7-branes. In the orientifold limit, $\tau$ is an arbitrary
constant, this $N=24$ solution is interpreted as modding out
$\text{Mink}_{1,7} \times T^2$, with compact transverse space,
with the same $\mathbb{Z}_2$-symmetry as before.

It was realized in \cite{Dasgupta:1996ij} that apart from the $N=6$ solution
and its embedding in the $N=24$ solution one could also consider
configurations with $N=24$ which are made by joining solutions whose
asymptotic value for $\tau$ is either $i$ or $\rho$. The asymptotic value of
$\tau$ for $N=6$ can be both $i$ and $\rho$, for the $N=9$ solution it must be
$i$, and for the $N=8,10$ solutions it must be $\rho$. If we take the orbifold
limit of the $N=6,8,9,10$ solutions we get the orbifolds points of
$T^4/\mathbb{Z}_{n}$ for $n=3,4,6$. For example the solution of
\cite{Dasgupta:1996ij} which in the orbifold limit $\tau=i$ becomes
$T^4/\mathbb{Z}_{4}$ is made out of two $N=9$ solutions and one $N=6$
solution.

Since the $N=6,12,24$ solutions all have a perturbative asymptotic
regime it is interesting to count the number of free parameters in
these solutions which correspond to the relative motion of the
D7-branes. In subsection \ref{blocks} it was derived that for
general $N$ the number of real free parameters is $2(2N-2)+1$. For
$N=6,12,24$ there are a number

\begin{equation}
    2(N-N/2+N-N/3)
\end{equation}

\noindent more real parameters as compared to the situation in
which all point $z_i$ are grouped in doublets and all points
$z_{\rho}$ are grouped in triplets. For example $N-N/3$ is the
number of parameters one has to fix to go from a configuration in
which non of the points $z_{\rho}$ are coincident to a
configuration in which all points $z_{\rho}$ are organized in
triplets. We count

\begin{equation}\label{numberfreeparameters}
    4N-3-2(N-N/2+N-N/3)=5N/3-3
\end{equation}

\noindent free real parameters. Thus for the $N=6,12,24$ F-theory
solutions there is one real parameter plus $3,8,18$ complex
parameters, respectively.

It has been argued in \cite{Vafa:1996xn} that the $N=24$
configuration consisting only of D7-branes can be interpreted as
an 8-dimensional supergravity coupled to 19 vector multiplets
obtained via a $K3$ compactification of F-theory\footnote{Of these
19 vector multiplets 18 are to
  attributed the presence of the 24 D7-branes (see equation \eqref{numberfreeparameters})
  and one comes from the reduction of
  the IIB supergravity (see footnote \ref{dividingbySL2Z}).  Note that there is an additional graviphoton sitting
  in the supergravity multiplet leading to a total of 20 vectors.}. In the
orientifold limit this theory has been argued to be dual to heterotic on a
two-torus \cite{Vafa:1996xn,Dabholkar:1997zd}.


\subsection{A solution with monodromy group $\Gamma_{0}(2)$}\label{subgroup}

Here we present an explicit realization of a solution whose
monodromy group is $\Gamma_{0}(2)$. We take for $\tau$ and $f$ the
following:

\begin{align}
  j_{\Gamma_{0}(2)}(\tau) &
=\frac{1}{(1+i)^{12}}\left(\frac{\eta(\tau)}{\eta(2\tau)}\right)^{24}
=\frac{z-z_0}{z-z_{i\infty}}\,, \\
& \nonumber \\
f &
=\eta(\tau)\eta(2\tau)(z-z_{0})^{-3/24}(z-z_{i\infty})^{-3/24}\,.
\end{align}

\noindent This solution consists of one D7-brane ($p=1$) and the
S-dual of a D7-brane with $T^2$ monodromy and no
$\text{det}\,Q>0$-branes. The monodromy of $\tau$ for this S-dual
D7-brane is $ST^2S$ ($q=2$). Within the subgroup $\Gamma_{0}(2)$
these two branes are inequivalent.

This simple example is included to show explicitly that one can
construct globally well--defined supersymmetric solutions with
finite energy without using the full $SL(2,\mathbb{Z})$ duality
group of type IIB and that a priori the monodromy group of a
7-brane configuration need not equal the full duality group. It it
clear that there is a wealth of other possible choices, leading to
different supersymmetric configurations of 7-branes. Here we will
not pursue this point any further.


\section{Conclusions}\label{sec:conclusions}

In this paper we have re-derived old results and found new
possibilities for globally well--defined, supersymmetric 7-brane
solutions of finite energy in type IIB supergravity. A number of
things are worth pointing out.

One of our findings is that the most general supersymmetric
solutions contain objects with $\det Q$ both equal to zero and
positive. By considering particular configurations of the
$N=6,12,24$ solutions the results of \cite{Greene:1989ya} were
reproduced, using a different and arguably more explicit argument.
In this paper the global properties of the Killing spinor have
played a key role, whereas \cite{Greene:1989ya} focussed on the
uplift to a manifold of special holonomy.

It is instructive to consider why we have found more
supersymmetric 7-brane configurations. The defining property of
the configurations of \cite{Greene:1989ya} is that they all lead
to smooth CY 2-folds in the $(z, \tau)$ directions. The
distinguishing feature of the new supersymmetric 7-brane
configurations we constructed in this paper is that they do not
correspond to a CY 2-fold reduction of F-theory.  This is due to
the presence of the $\det\, Q>0$-branes which lead to conical
singularities at the points $z_i$ and $z_\rho$ that cannot be
resolved into a Ricci--flat K\"ahler manifold. It would, however,
still be a manifold of $SU(2)$ holonomy.

The status and role of the $\det Q >0$-branes in string theory is
at present unclear. One can adopt a number of different
viewpoints. For example, an interpretation as bound states of
D7-branes and $SL(2,\mathbb{Z})$ transforms thereof can be put
forward since any $\det Q
>0$ monodromy can be written as the product of $\det Q =0$ monodromies.
Indeed, the approximate solutions \eqref{taudetq>0},
\eqref{taudetq=0} around the points $z_\rho$ and $z_i$ can be
written in terms of a distribution of D7-branes
\cite{Bergshoeff:2004nq}. Similar bound states occurred when we
discussed solutions with $\Gamma_0(2)$ monodromy group in
subsection \ref{subgroup}.

Another interpretation is in terms of O7-planes, as has been
discussed in section \ref{D7-trunc}. The results of the present
paper are more in line with the latter point of view. The D7-brane
necessarily comes with two additional points $z_i$ and $z_\rho$,
to cancel its mass and $SL(2,\mathbb{Z})$-charge, and thus to
allow for a globally well--defined solution. This is reminiscent
of the D8-brane, which also is necessarily paired with an
O8-plane, see e.g.~\cite{Polchinski:1995df, Bergshoeff:2001pv}.
The difference is that the D8-brane carries Abelian charge, which
is exactly cancelled by the O8-plane, while the D7-brane requires
points of $SL(2,\mathbb{Z})$ monodromy. Of course, for $p$-branes
with $p \leq 6$ these complications do not occur since their
transverse space is at least 3-dimensional, and hence allows for a
net charge.

It turns out that there is an intriguing relation between the
approximate $D=10$ 7-brane solutions
\eqref{taudetq=0}-\eqref{taudetq>0}, which are expansions of
modular functions around the orbifold points of the $\tau$-plane,
and domain walls in $D=9$ dimensions. In fact, the approximate
7-brane solutions for $\det\, Q >0$ were first found by uplifting
domain walls of $D=9$ gauged maximal supergravity with gauge group
$SO(2)$ \cite{Bergshoeff:2002mb}. The reason for this relationship
is that the approximate 7-brane solutions, in contrast to the full
solution, have an isometry in their transverse space. More
precisely, their angular dependence (in the transverse space) is
of the form of an $SL(2,\mathbb{Z})$ transformation and this
allows a twisted Scherk--Schwarz reduction over this direction to
$D=9$ dimensions.

In general, the supersymmetric domain walls of gauged maximal
supergravities in $D$ dimensions give rise to the near-horizon
limit of (distributions of) $p$-branes with $D=p+2$
\cite{Boonstra:1998mp, Bergshoeff:2004nq}. For 7-branes, however,
there are problems with the definition of the near-horizon limit
(due to the absence of a corresponding `dual' frame). In view of
the domain wall connection, it thus seems that the expansions
\eqref{taudetq=0}, \eqref{taudetq>0} around the orbifold points
should be seen as the analogue of near-horizon limits for
7-branes. Indeed, in this limit there is an $S^1$ isometry in the
solution, in analogy with the spherical part in the near-horizon
limit of $p$-branes with $p \leq 6$. It would be interesting to
pursue these ideas further.


\section*{Acknowledgements}

We would like to thank D.~Sorokin for useful discussions, in
particular, regarding section \ref{sourceterms}. T.O.~and
D.R.~would like to thank the University of Groningen for
hospitality, while J.H.~and D.R.~would like to thank the
Universidad Aut\'{o}noma in Madrid for hospitality. E.B.~and
T.O.~are supported by the European Commission FP6 program
MRTN-CT-2004-005104 in which E.B.~is associated to Utrecht
university and T.O.~is associated to the IFT-UAM/CSIC in Madrid.
The work of E.B.~and T.O.~is partially supported by the Spanish
grant BFM2003-01090. The work of T.O.~has been partially supported
by the Comunidad de Madrid grant HEPHACOS P-ESP-00346. Part of
this work was completed while D.R.~was a post-doc at King's
College London, for which he would like to acknowledge the PPARC
grant PPA/G/O/2002/00475. In addition, he is presently supported
by the European EC-RTN project MRTN-CT-2004-005104, MCYT FPA
2004-04582-C02-01 and CIRIT GC 2005SGR-00564. J.H. is supported by
a Breedte Strategie grant of the University of Groningen.


\end{document}